\documentclass{emulateapj}
\usepackage{lscape}

\slugcomment{Accepted for publication in the {\it Astrophysical Journal}}

\shorttitle{Fe K Line Profile in PG Quasars}
\shortauthors{Inoue et al.}

\begin{document}


\title{Fe K Line Profile in Low-redshift Quasars: \\Average Shape and Eddington Ratio Dependence}

\author{Hirohiko  Inoue\altaffilmark{1}}
\affil{Institute of Space and Astronautical Science, 3-1-1 Yoshinodai, Sagamihara, Kanagawa 229-8510, Japan}
\email{hirohiko@astro.isas.jaxa.jp}

\author{Yuichi  Terashima}
\affil{Department of Physics, Ehime Univesity, 2-5 Bunkyo-cho, Matsuyama, Ehime 790-8577, Japan}

\author{Luis C. Ho\altaffilmark{}}
\affil{The Observatories of the Carnegie Institution of Washington, 813 Santa Barbara Street, Pasadena, CA 91101-1292}

\altaffiltext{1}{Department of Physics, Tokyo Institute of Technology, 2-12-1 
Ohokayama, Meguro, Tokyo 152-8551, Japan}

\begin{abstract}

We analyze X-ray spectra of 43 Palomar-Green quasars observed with
{\it XMM-Newton} in order to investigate their mean Fe K line profile
and its dependence on physical properties. The continuum spectra of 39
objects are well reproduced by a model consisting of a power law and a
blackbody modified by Galactic absorption.  The spectra of the
remaining four objects require an additional power-law component
absorbed with a column density of $\sim 10^{23} {\rm cm}^{-2}$.  A
feature resembling an emission line at 6.4 keV, identified with an Fe
K line, is detected in 33 objects.  Approximately half of the sample
show an absorption feature around 0.65--0.95 keV, which is due to
absorption lines and edges of \ion{O}{7} and \ion{O}{8}. We fit
the entire sample simultaneously to derive average Fe line parameters
by assuming a common Fe line shape. The Fe line is relatively narrow
($\sigma=0.36$\,keV), with a center energy of 6.48\,keV and a mean
equivalent width (EW) of 248\,eV.  By combining black hole masses
estimated from the virial method and bolometric luminosities derived
from full spectral energy distributions, we examine the dependence of
the Fe K line profile on Eddington ratio. As the Eddington ratio
increases, the line becomes systematically stronger (EW = 130 to 280
eV), broader ($\sigma=0.1$ to 0.7\,keV), and peaks at higher energies
(6.4 to 6.8\,keV).  This result suggests that the accretion rate onto
the black hole directly influences the geometrical structure and
ionization state of the accretion disk.  We also examine a
two-component model consisting of a Gaussian and a diskline to
constrain the intensity of the broad line. The mean equivalent widths
are $\approx 70-180$\,eV for the four Eddington ratio groups, although
the standard deviations in each group are very large. This suggests
that the broad line is not ubiquitous.

\end{abstract}


\section{Introduction}

Fluorescent Fe K lines are common in the X-ray spectra of accreting black 
holes, ranging from X-ray binaries to active galactic nuclei (AGNs), and 
provide one of the best probes for studying accretion disks. If the Fe K line 
is emitted from the inner part of the accretion disk, it becomes broad and 
asymmetric due to both Doppler shift and gravitational redshift (Fabian et al. 
1989). The best example of a broad Fe K line is seen in the Seyfert 1 galaxy 
MCG--6--30--15 (Tanaka et al. 1995). Broad Fe K lines have been observed in 
many other sources and are thought to be very common in the X-ray spectra of
Seyfert 1 nuclei (Nandra et al. 1997a).

Recent observations of AGNs with {\it XMM-Newton} show that the
broad Fe K line is not as common as previously believed.  While a few
Seyfert 1 galaxies (e.g., MCG--6--30--15, Mrk 205, Mrk 509) indeed show an 
unambiguous broad line, Fe lines in other galaxies are dominated by a 
relatively narrow feature.  From an analysis of 53 type 1 AGNs observed with 
{\it XMM-Newton}, 
 Page et al. (2004a) showed that a broad Fe K line ($\sigma \geq
0.1$\,keV) is seen in only 13 sources, while the remaining 40
sources have a narrow line ($\sigma \leq 0.1$\,keV) or no lines. Some of
the 13 sources show a narrow line as well as the broad line.
In a similar study, 
Yaqoob \& Padmanabhan (2004) reported the results of 18 observations of 15 
Seyfert 1 galaxies observed with the {\it Chandra} High Energy Grating. They 
measured the width of the line core and obtained a weighted mean of FWHM = 
2380$\pm$ 760 km\,s$^{-1}$, which is slightly larger than the instrument 
resolution (FWHM $\approx$ 1860 km\,s$^{-1}$). Evidence of underlying 
broad-line emission was also found in at least four sources.   Based on these 
recent studies, it is still controversial whether or not relativistically 
broadened Fe K emission is truly common in nearby AGNs.

Streblyanska et al. (2005) derived an average rest-frame spectrum of
AGNs detected in a 770 ksec {\it XMM-Newton}
observation of the Lockman Hole field. They used a sample of 104 X-ray
sources with optical redshifts measured by Lehmann et al. (2001), analyzing
separately the type 1 and type 2 subsamples defined by optical spectra
(Schmidt et al. 1998; Lehman et al. 2001). 
From composite rest-frame spectra generated for the type 1 and type 2 sources, 
they found evidence for a broad line peaking at a rest-frame energy 
$\sim$6.4\,keV with an equivalent width (EW) $\sim$560\,eV and $\sim$460\,eV, 
respectively.  However, it should be noted that there are some systematic 
uncertainties in modelling the
continuum, since it is virtually impossible to distinguish a very
broad line clearly from the continuum for sources as faint as those
analyzed by Streblyanska et al. (2005).  In addition, making
composite spectra of faint sources may introduce spurious spectral
features (Yaqoob 2005).

Brusa et al. (2005) and Comastri et al. (2007) studied average
spectra of AGNs detected in the {\it Chandra} Deep Field North and
South.  They stacked the X-ray counts in the observed frame from
spectroscopically identified AGNs, using a 
large number of source spectra within sufficiently narrow redshift ranges 
such that the energy (redshift) spread is negligible. A broad
Fe line is seen in some of the stacked spectra and is fitted with a
relativistic disk line model peaking at a rest-energy of 6.4
keV. Since the fluxes of the objects in the sample are low, this study 
may suffer from the same uncertainties affecting the Streblyanska et al. 
analysis of the Lockman Hole.

What parameters of the AGN affect the strength or profile of the Fe K line?
An inverse correlation between the equivalent width of Fe K line and X-ray 
luminosity, known as the X-ray Baldwin effect, was first pointed out by 
Iwasawa \& Taniguchi (1993).  This correlation has been confirmed with 
{\it ASCA} (Nandra et al. 1997b) and {\it XMM-Newton} data (Page et
al. 2004a; see also Jim\'enez-Bail\'on et al. 2005).  
The dependence of the iron line profile with luminosity was
investigated by Nandra et al. (1997b) using 18
Seyfert 1s and 21 quasars observed with {\it ASCA}. They divided the
sample into five luminosity bins and examined average Fe K line
profiles. The line shows a very similar profile, which is composed of
a narrow core and a broad red wing for groups with a luminosity below
$L_{\rm X} \approx 10^{44}\,{\rm ergs~s^{-1}}$. The intensity of the
narrow core, however, becomes weak above this luminosity. The red wing
becomes weak and the peak energy shifts to higher energy for the
luminosity range of $10^{45}\le L_{\rm X} \le 10^{46}\,{\rm ergs~
s^{-1}}$; no evidence for line emission is observed above $L_{\rm X}
\approx 10^{46}\,{\rm ergs~ s^{-1}}$.  Jim\'enez-Bail\'on et al. (2005), 
analyzing a sample of 39 Palomar-Green (PG; Schmidt \& Green 1983) quasars 
observed with {\it XMM-Newton}, found a similar result as that reported by 
Nandra et al. (1997b). The luminosity dependence of the Fe K line may be 
related to the ionization level of the accretion disk.

This paper examines the behavior of the Fe K line
as a function of accretion rate, as parameterized by 
the Eddington ratio, taking advantage of the availability of black hole mass 
and bolometric luminosity estimates.  The Eddington ratio is a fundamental 
parameter that may control both the ionization state and geometric structure 
of the accretion disk.  If so, we expect the profile of the Fe K line, 
in particular its width and central energy, to vary systematically with 
Eddington ratio.  The goal of this study is to investigate this issue.


This paper is organized as follows. Section 2 describes the
observations, and Section 3 shows the results of the spectral analysis of
average spectra and its Eddington ratio dependence. We discuss the
physical origin of the Eddington ratio dependence in Section 4. Our
findings are summarized in Section 5.

\section{{\it XMM-Newton} Observations}
\subsection{The Sample}

We selected our sample from the 87 low-redshift ($z<0.5$)
Palomar-Green quasars in Boroson \& Green (1992).  X-ray data for 45
objects are available in the {\it XMM-Newton}
Science Archive (XSA).  Only EPIC-PN data were used in the following
analysis since the effective area of PN is much larger than that of
EPIC-MOS.  PG 1001+045 was excluded from the present study because no
EPIC-PN data are available.  PG 1226+023 (3C 273), a very bright quasar,
was also excluded since it overwhelms the X-ray flux in the average
spectrum examined below.  Basic data (name, redshift, central black
hole mass, and Eddington ratio) for the final sample of 43 objects are
given in Table \ref{tab: the sample}. {\it XMM-Newton} results of 34
objects have been published, and the references are also given in Table
\ref{tab: the sample}. A log of the
observations is given in Table \ref{tab: the sample xray}.  

The black hole masses and Eddington ratios are taken from the study of 
L. C. Ho (2007, in preparation).  Ho calculates black hole masses employing 
the virial method of Kaspi et al. (2000), using broad H$\beta$ line widths 
from Boroson \& Green (1992) and 5100 \AA\ continuum luminosities from 
Neugebauer et al. (1987).  His bolometric luminosities come from direct 
integration of the full, broad-band (radio to X-rays) spectral energy 
distribution rather than from a single band assuming a bolometric correction.
All distance-dependent quantities were derived assuming the following 
cosmological parameters:  $\Omega_{\rm m}=0.3, \Omega_{\Lambda}=0.7$, and 
$H_0$ = 72 kms Mpc$^{-1}$.  Eddington luminosities, computed 
assuming a composition of pure hydrogen, are given by

\begin{equation}
        L_{\rm Edd}= \frac{4\pi GM_{\rm BH}m_{\rm p}c}{\sigma_{\rm T}} = 1.26\times
 10^{38}\left(\frac{M_{\rm BH}}{M_\odot}\right) {\rm ergs\,s^{-1}},
\end{equation}
where $M_{\rm BH}$ is the mass of the black hole, $m_{\rm p}$ is the proton 
mass, and $\sigma_{\rm T}$ is the Thompson scattering cross-section.  

\begin{figure*}[htbp]
\begin{center}
\includegraphics[angle=-90,scale=.29]{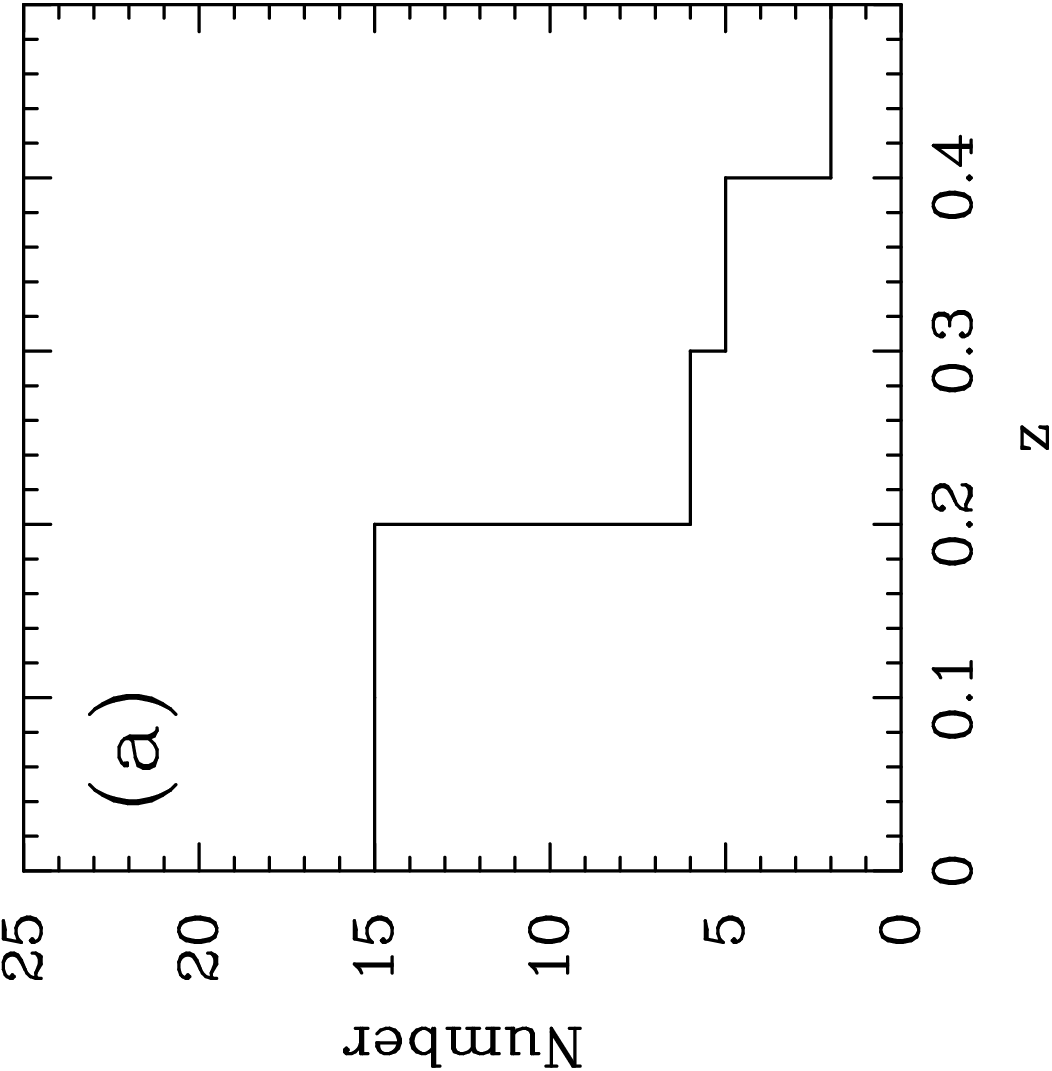}
\includegraphics[angle=-90,scale=.29]{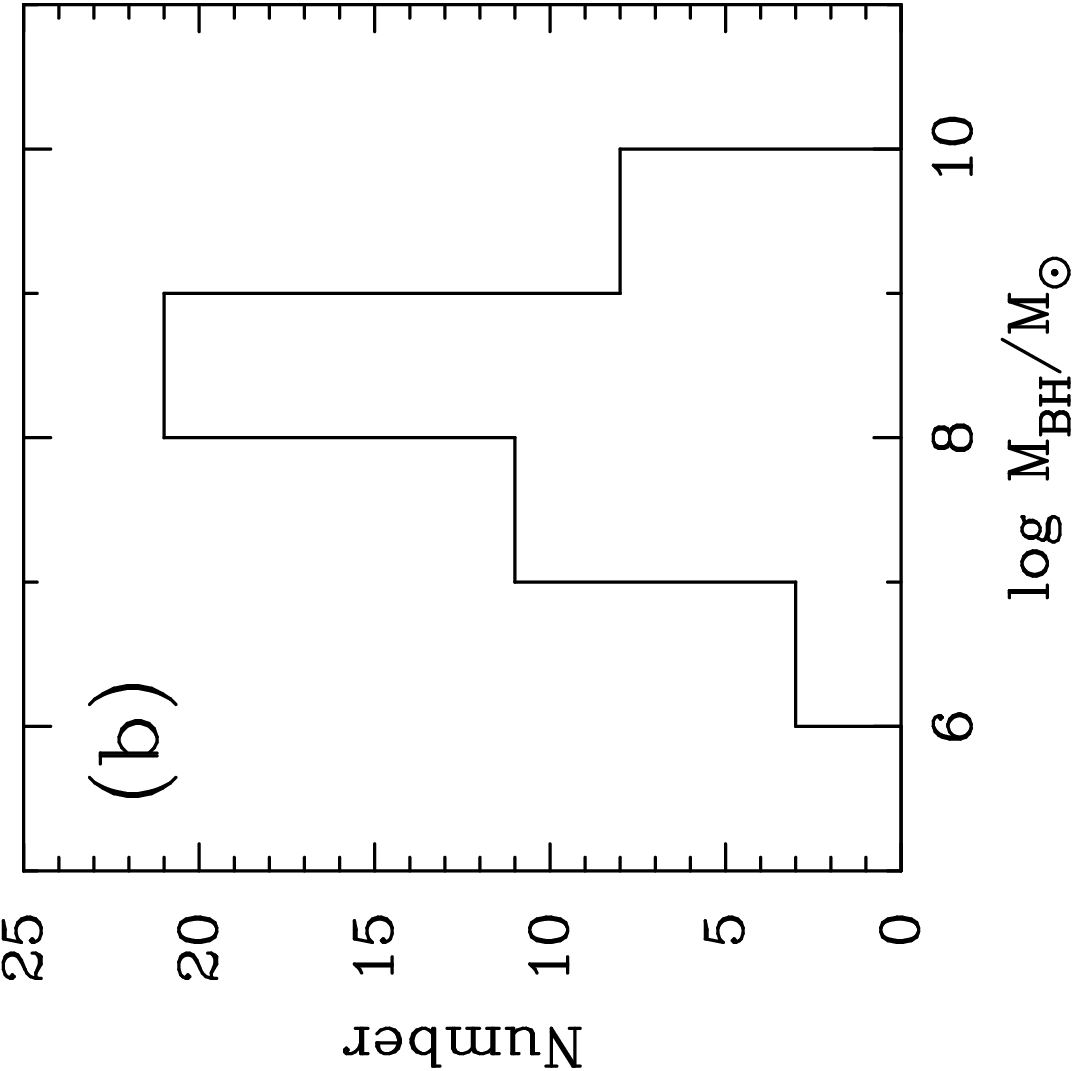}
\includegraphics[angle=-90,scale=.29]{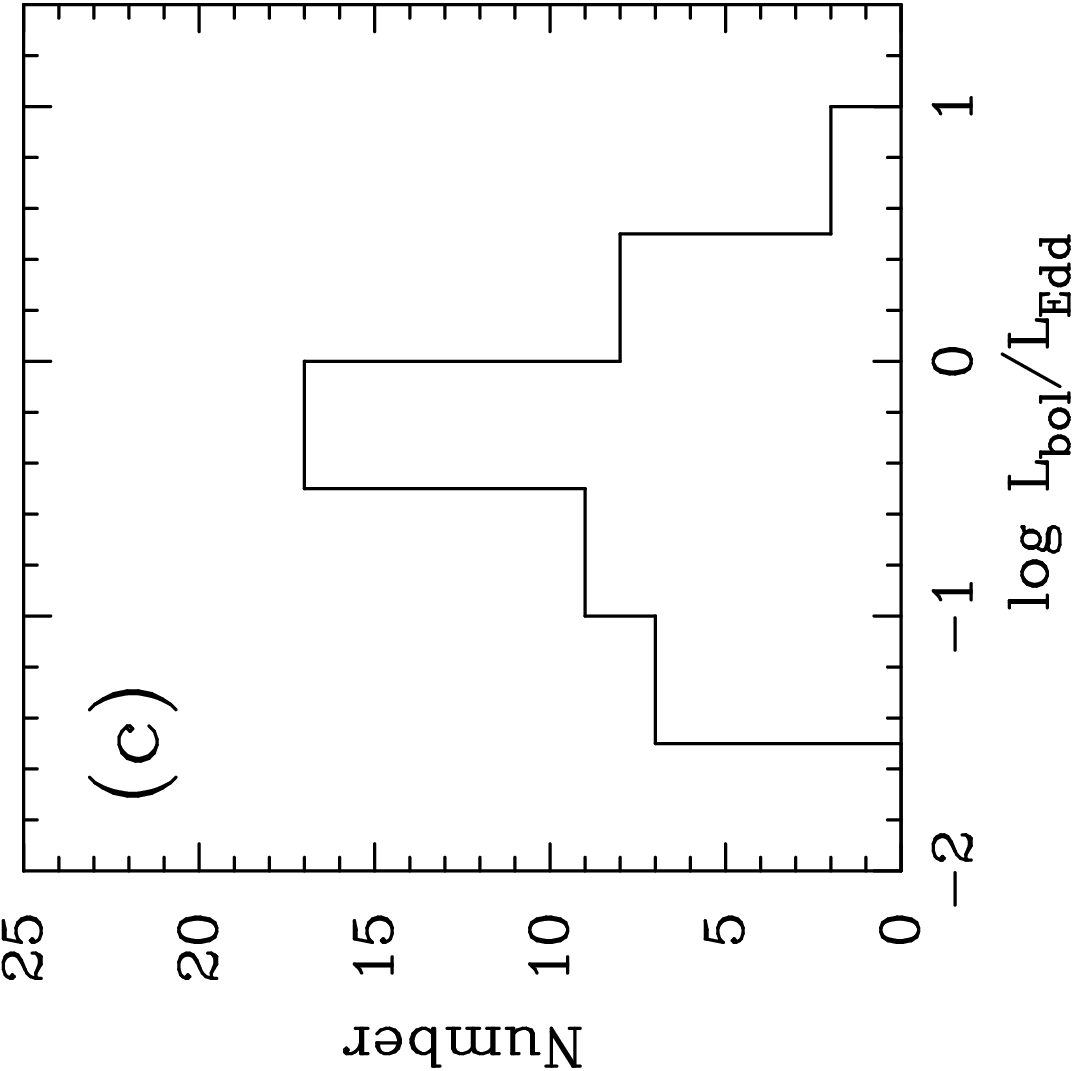}
\caption{Distribution of ({\it a}) redshifts, ({\it b}) 
central black hole masses, and ({\it c}) Eddington ratios.}
\label{fig: redshift mass edd_ratio dist}
\end{center}
\end{figure*}

Figure\,\ref{fig: redshift mass edd_ratio dist} shows the
distributions of redshift, black hole mass, and Eddington ratios.
The mean redshift is $z\approx0.2$, but the sample covers a wide range of 
black hole masses ($10^6 < M_{\rm BH}/M_\odot\ < 10^9$) and Eddington ratios 
($0.05 < L_{\rm bol}/L_{\rm Edd} < 4.5$).

\begin{deluxetable*}{cccccccl}
\tabletypesize{\scriptsize}
\tablecaption{The Sample\label{tab: the sample}}
\tablewidth{0pt}
\tablehead{
\colhead{PG } & \colhead{Alternate Name} & \colhead{$z$} &
 \colhead{log ($M_{\rm BH}/M_\odot$)} & \colhead{$L_{\rm bol}/L_{\rm Edd}$} &
\colhead{Group\tablenotemark{a}} & \colhead{Note\tablenotemark{b}} & \colhead{Reference} \\
} 
\startdata
0003+199 & Mrk 335 & 0.026 & 7.16  & 0.58  &2  &           & 1, 2\\ 
0007+106 & Mrk 1501& 0.089 & 8.69  & 0.10  &4  & RL    &3, 4 \\ 
0050+124 & I Zw 1  & 0.061 & 7.29  & 2.51  &1  &           &2, 3, 4, 5\\ 
0157+001 & Mrk 1014& 0.163 & 8.21  & 1.11  &1  &           &2, 3, 4 \\ 
0804+761 & ...     & 0.100 & 8.41  & 0.45  &2  &           &3, 4, 5, 6 \\ 
0844+349 & Ton 951 & 0.064 & 7.80  & 0.33  &3  &           &2, 3, 4, 7, 8, 9\\ 
0947+396 & ...     & 0.206 & 8.67  & 0.26  &3  &           &2, 3, 4, 5, 10 \\ 
0953+414 & ...     & 0.234 & 8.69  & 0.66  &2  &           &2, 3, 4, 5, 10 \\ 
1004+130 & PKS 1004+13 & 0.240 & 9.43 & 0.09 & 4 &RL, BAL  & 11\\ 
1048+342 & 3C 246  & 0.167 & 8.36  & 0.35  &3  &           &2, 3, 4, 5, 10 \\ 
1100+772 & 3C 249.1& 0.312 & 9.43  & 0.12  &4  &RL &3, 4 \\ 
1114+445 & ...     & 0.144 & 8.57  & 0.13  &4  &           &3, 4, 5, 10, 12 \\ 
1115+407 & ...     & 0.154 & 7.61  & 1.91  &1  &           &2, 3, 4, 5, 10 \\ 
1116+215 & Ton 1388& 0.177 & 8.64  & 0.68  &2  &           &2, 3, 4, 5, 10 \\
1202+281 & GQ Comae& 0.165 & 8.54  & 0.18  &3  &           &2, 3, 4, 5, 10\\ 
1211+143 & ...     & 0.081 & 7.95  & 0.79  &2  &           &2, 3, 4, 13, 14\\ 
1216+069 & ...     & 0.331 & 9.40  & 0.18  &3  &           &3, 4, 10\\ 
1244+026 & ...     & 0.048 & 6.39  & 4.57  &1  &           &2, 3, 4, 5 \\ 
1302$-$102 & PKS 1302$-$102& 0.27  & 9.04  & 0.48  &2  &RL & \\ 
1307+085 & ...     & 0.155 & 8.87  & 0.10  &4  &           &2, 3, 4, 5 \\ 
1309+355 & Ton 1565& 0.184 & 8.37  & 0.32  &3  &RL &2, 3, 4, 5, 10, 12 \\ 
1322+659 & ...     & 0.168 & 8.22  & 0.49  &2  &           &2, 3, 4, 5, 10 \\ 
1352+183 & ...     & 0.152 & 8.37  & 0.37  &3  &           &2, 3, 4, 5, 10 \\ 
1402+261 & Ton 182 & 0.164 & 7.94  & 1.78  &1  &           &2, 3, 4, 5, 10, 15\\ 
1404+226 & ...     & 0.098 & 6.80  & 2.63  &1  &           &2, 3, 4, 16, 17 \\ 
1411+442 & ...     & 0.090 & 7.98  & 0.28  &3  &        &3, 4, 10, 18 \\ 
1415+451 & ...     & 0.114 & 7.87  & 0.38  &3  &           &3, 4, 10 \\ 
1425+267 & Ton 202 & 0.366 & 9.75  & 0.05  &4  & RL        &10, 19 \\ 
1426+015 & Mrk 1383& 0.087 & 8.98  & 0.08  &4  &           &5, 6 \\ 
1427+480 & ...     & 0.221 & 8.10  & 0.71  &2  &           &2, 3, 4, 5, 10 \\ 
1440+356 & Mrk 478 & 0.079 & 7.42  & 1.91  &1  &           &2, 3, 4, 5, 10 \\ 
1444+407 & ...     & 0.267 & 8.36  & 0.94  &2  &           &2, 3, 4, 10 \\ 
1448+273 & ...     & 0.065 & 6.95  & 1.86  &1  &           & \\ 
1501+106 & Mrk 841 & 0.036 & 8.39  & 0.07  &4  &           &2, 3, 4 \\ 
1512+370 & 4C +37.43 & 0.371 & 9.50  & 0.13  &4& RL        &5, 10 \\ 
1535+547 & Mrk 486 & 0.039 & 7.02  & 0.41  &3  &        &20\\ 
1543+489 & ...     & 0.400 & 8.13  & 3.39  &1  &           &10 \\ 
1613+658 & Mrk 876 & 0.129 & 9.12  & 0.06  &4  &           &3, 4, 5, 6 \\ 
1626+554 & ...     & 0.133 & 8.43  & 0.19  &3  &           &3, 4, 5, 10\\ 
2112+059 & ...     & 0.466 & 9.27  & 0.56  &2  &BAL        &20\\ 
2130+099 & Mrk 1513& 0.063 & 7.82  & 0.65  &2  &           & \\ 
2214+139 & Mrk 304 & 0.066 & 8.38  & 0.06  &4  &           &3, 4, 18 \\ 
2233+134 & ...     & 0.325 & 8.15  & 2.29  &1  &           & \\ 
\enddata
\tablerefs{
(1) Gondoin et al. 2002, 
(2) Crummy et al. 2006, 
(3) Piconcelli et al. 2005, 
(4) Jim\'enez-Bail\'on et al. 2005, 
(5) Porquet et al. 2004, 
(6) Page et al. 2004b, 
(7) Brinkmann et al. 2006, 
(8) Brinkmann et al. 2003, 
(9) Pounds et al. 2003, 
(10) Brocksopp et al. 2006, 
(11) Miller et al. 2006, 
(12) Ashton  et al. 2004, 
(13) Pounds et al. 2003, 
(14) Pounds \& Page 2006,  
(15) Reeves et al. 2004, 
(16) Crummy et al. 2005, 
(17) Dasgupta et al. 2005, 
(18) Brinkmann et al. 2004, 
(19) Miniutti \& Fabian 2006, 
(20) Schartel et al. 2005.
}
\tablenotetext{a}{Group divided by Eddington ratio:  (1)\,$1.00 \le L_{\rm bol}/L_{\rm Edd}$, (2)\,$0.40 \le
L_{\rm bol}/L_{\rm Edd}\le 1.00$, (3)\,$0.17 \le L_{\rm bol}/L_{\rm
Edd}\le 0.40$, (4)\,$L_{\rm bol}/L_{\rm Edd} \le 0.17$. }
\tablenotetext{b}{RL: Radio-loud quasars (Boroson \& Green
 1992), BAL: Broad absorption-line quasars (Weymann et al. 1991; Brandt et al. 2000).}
\end{deluxetable*}


\begin{deluxetable*}{ccccccc}
\tabletypesize{\scriptsize}
\tablecaption{Observation Log
\label{tab: the sample xray}}
\tablewidth{0pt}
\tablehead{
\colhead{PG } & \colhead{Obs. Date} & \colhead{Exposure} &
 \colhead{Count rate\tablenotemark{a}} & \colhead{$F_{\rm 2-10\,keV}$
\tablenotemark{b}} &
\colhead{$L_{\rm 2-10\,keV}$\tablenotemark{c}} & \colhead{Note\tablenotemark{d}} \\
\colhead{ } & \colhead{ } & \colhead{(ks)} &
 \colhead{(count\,s$^{-1}$)} & \colhead{($10^{-12}{\rm ergs\,s^{-1}\,cm^{-2}}$)} &
\colhead{($10^{44}{\rm ergs\,s^{-1}}$)} & \colhead{}
} 
\startdata
0003+199 & 2000 Dec 25 & 28.5 & 19.45  & 13.75   & 0.20     & 1 \\ 
0007+106 & 2000 Jul 03 & 10.2 & 3.23   & 7.01    & 1.28     & \\  
0050+124 & 2002 Jun 22 & 18.3 & 7.95   & 8.17    & 0.71     & 1 \\ 
0157+001 & 2000 Jul 29 & 4.4  & 1.10   & 0.91    & 0.65     & \\ 
0804+761 & 2000 Nov 04 & 0.5  & 11.22  & 9.94    & 2.45     & \\ 
0844+349 & 2000 Nov 04 & 10.3 & 6.45   & 4.87    & 0.46     & \\ 
0947+396 & 2001 Nov 03 & 17.6 & 1.76   & 1.78    & 2.05     & \\ 
0953+414 & 2001 Nov 22 & 10.8 & 3.92   & 2.94    & 4.65     & 1 \\ 
1004+130 & 2003 May 04 & 18.2 & 0.10   & 0.32    & 0.48     & \\ 
1048+342 & 2002 May 13 & 21.5 & 1.17   & 1.30    & 0.94     & \\ 
1100+772 & 2002 Nov 01 & 6.8  & 2.02   & 3.62    & 10.16   & \\ 
1114+445 & 2002 May 14 & 35.0 & 0.53   & 2.28    & 1.12     & 1 \\ 
1115+407 & 2002 May 17 & 14.4 & 2.20   & 1.19    & 0.75     & \\ 
1116+215 & 2001 Dec 02 & 5.5  & 5.05   & 3.19    & 2.77     & \\ 
1202+281 & 2002 May 30 & 12.2 & 2.64   & 3.58    & 2.46     & 1 \\ 
1211+143 & 2001 Jun 15 & 48.5 & 3.33   & 3.05    & 0.46     & 2 \\ 
1216+069 & 2002 Jul 05 & 5.4  & 0.82   & 1.12    & 3.63     & \\ 
1244+026 & 2001 Jun 17 & 4.6  & 6.15   & 2.31    & 0.12     & \\ 
1302$-$102&2002 Jan 03 & 11.5 & 0.61   & 3.05    & 6.44     & \\ 
1307+085 & 2002 Jun 13 & 9.9  & 0.80   & 1.96    & 1.13     & 1 \\ 
1309+355 & 2002 Jun 10 & 23.6 & 0.44   & 0.70    & 0.61     & 1 \\ 
1322+659 & 2002 May 11 & 8.0  & 2.41   & 1.27    & 0.98     & 1 \\ 
1352+183 & 2002 Jul 20 & 8.9  & 2.30   & 1.87    & 1.11     & 1 \\ 
1402+261 & 2002 Jan 27 & 9.1  & 3.17   & 1.80    & 1.32     & \\ 
1404+226 & 2001 Jun 18 & 9.1  & 0.59   & 0.13    & 0.03     & 2 \\ 
1411+442 & 2002 Jul 10 & 21.6 & 0.12   & 0.46    & 0.08     & 1 \\ 
1415+451 & 2002 Dec 08 & 20.4 & 1.43   & 1.06    & 0.34     & \\ 
1425+267 & 2002 Jul 28 & 29.6 & 0.67   & 1.68    & 6.29     & 1 \\ 
1426+015 & 2000 Jul 28 & 2.2  & 9.39   & 8.48    & 1.50     & \\ 
1427+480 & 2002 May 31 & 32.5 & 1.15   & 1.05    & 1.44     & \\ 
1440+356 & 2003 Jan 04 & 10.8 & 5.58   & 2.63    & 0.39     & \\ 
1444+407 & 2002 Aug 11 & 16.7 & 0.96   & 0.45    & 1.05     & \\ 
1448+273 & 2003 Feb 08 & 18.0 & 4.23   & 2.07    & 0.20     & \\ 
1501+106 & 2001 Jan 13 & 7.6  & 18.31  & 15.00   & 0.44     & 1 \\ 
1512+370 & 2002 Aug 25 & 15.4 & 1.54   & 1.87    & 7.98     & 1 \\ 
1535+547 & 2002 Nov 03 & 14.4 & 0.16   & 1.33    & 0.04     & \\ 
1543+489 & 2003 Feb 08 & 3.7  & 0.35   & 0.13    & 0.87     & \\ 
1613+658 & 2001 Aug 29 & 2.4  & 4.46   & 4.98    & 2.05     & 1 \\ 
1626+554 & 2002 May 05 & 4.8  & 2.84   & 2.86    & 1.28     & \\ 
2112+059 & 2003 May 14 & 7.4  & 0.13   & 0.34    & 2.24     & \\ 
2130+099 & 2003 May 16 & 23.3 & 2.13   & 3.63    & 0.32     & 1 \\ 
2214+139 & 2002 May 12 & 6.2  & 0.55   & 3.32    & 0.31     & 1 \\ 
2233+134 & 2003 May 28 & 6.2  & 0.59   & 0.49    & 1.69     & 1 \\
\enddata
\tablenotetext{a}{~Count rate in 0.3--15 keV.}
\tablenotetext{b}{~Flux in 2--10 keV, not corrected for absorption.}
\tablenotetext{c}{~Luminosity in 2--10 keV, not corrected for absorption.}
\tablenotetext{d}{~Signature of ionized absorber. 1: one edge; 2: two edges.}
\end{deluxetable*}

\vspace{0.5cm}

\subsection{Data Reduction}

Raw data from the {\it EPIC-PN} instrument were reprocessed using
Science Analysis Software (SAS) version 6.0 to produce calibrated
event files.  Only events with the single or double pixels (PATTERN
$\le$4) were used in the following analysis. Time intervals of
background flaring were excluded. Source spectra were extracted from a
circular region centered at the peak of the X-ray source
counts. Background spectra were taken from an off-source region in the
same field-of-view and subtracted from the source spectra.
Appropriate response and ancillary files were created using the RMFGEN
and ARFGEN tools in SAS, respectively. All the spectra were binned
so that each bin contains at least 30 (or more for bright sources)
counts in order to apply a $\chi^2$ minimization technique. Spectral
fits were performed with {\it XSPEC} version\,12.2.


\section{Results}

\subsection{Spectra of Individual Objects}

The spectra of most objects are not well represented by a simple power law; 
soft excess emission and an emission line-like feature around 6 keV are also
often seen. Therefore, we considered a model consisting of a power law, a
blackbody component to represent the soft excess emission below $\sim$ 2\,keV, 
and a Gaussian emission line around 6 keV, all modified by Galactic
absorption. We used a two blackbody model for the soft excess
if the single blackbody did not represent the data well. The hydrogen
column densities of the Galactic absorption are left free. Some objects show absorption features around 0.65--0.95 keV. We 
multiplied one or two edges to the above model for such cases.

The spectra of 39 objects are well reproduced by this model. We
obtained hydrogen column densities of $N_{\rm H}$ $\approx$
(0.003$-$0.26)$\times 10^{22}{\rm cm^{-2}}$, photon indices $\Gamma
\approx$ 1.5 to 2.5, and temperatures of the blackbody 
$kT\approx 0.07-0.23$\,keV.  The column densities are consistent with
the Galactic value derived from the HI map by Dickey \& Lockman (1990).
Except for a few cases (see Appendix), our 
results for individual objects are in
good agreement with the independent analysis published in the references given
in Table 1.  The spectra of the
remaining four objects (PG 1004+130, 1411+442, 1535+547, and
2214+139) show broad excess emission at energies above $\sim$ 2--3
keV. We added an absorbed power-law component to the model used above
to represent this component. This model provides a good description of
the spectra. The hydrogen column densities of the absorbed power law
are $\sim 10^{23}\,{\rm cm^{-2}}$ for all the cases. Details of the
spectral fits to these four quasars are given in Appendix\,\ref{appe:
pg qso}. The spectral parameters for the other components are similar
to those obtained for the 39 objects without an additional absorbed
power law.

An absorption feature around 0.65--0.95 keV, which can be well fitted by 
one or two edges, is seen in 20 out of the 43 
objects.  The majority of the sources (18/20) require only a single
edge, with a mean edge energy of $E=0.69$\,keV and an
optical depth of $\tau=0.56$.  The absorption features in the two remaining 
objects are represented by two edges:  for PG 1211+143, the edge energies are
$E=0.75^{+0.02}_{-0.01}$/$0.93^{+0.02}_{-0.01}$\,keV  and the optical depths
are $\tau=0.36^{+0.05}_{-0.05}$/$0.31^{+0.04}_{-0.07}$; for 
 PG 1404+226, the prameters are
$E=0.74^{+0.12}_{-0.06}$/$0.94^{+0.04}_{-0.03}$\,keV and
$\tau=0.25^{+0.22}_{-0.21}$/$0.89^{+0.38}_{-0.37}$).
Similar features are often seen in Seyfert 1s
and quasars observed with the energy resolution of CCD. High-resolution 
spectroscopy of such features with grating spectrometers
onboard {\it XMM-Newton} and {\it Chandra} have shown that they 
consist of many absorption lines of ionized species such as He-like
and hydrogenic O, Ne, and Mg.  The observed edge energies are in
agreement with such ionized absorbers observed with CCD resolution
(e.g., George et al. 1998, 2000).

The fluxes and luminosities obtained for the best-fit model,
as well as the detected count rates, are summarized in Table 2. 
Neither the fluxes nor the luminosities are corrected for absorption.

\subsection{Co-added Spectra}\label{sub: adding way}

We derived a co-added rest-frame spectrum of the 43 quasars in order
to examine the average shape of the Fe K emission line. Since each object
has a different redshift and thus a different observed line energy, the
detector response, which is energy dependent, is different for each
object. As the co-added rest-frame spectrum is a summation of spectra
taken with different energy resolution, this situation complicates
quantitative analysis. Therefore, the spectrum derived here is for
presentation purpose only, and quantitative analysis by simultaneous
fitting is done in the next subsection. 

The co-added spectrum was constructed as follows.  First, ratios of
the data to the best-fit continuum model were calculated.  The ratios
were then multiplied by the unfolded best-fit continuum model.  The
resulting spectrum is composed of an unfolded continuum and a folded
emission-line component.  The energy scales of both the unfolded
continuum spectrum and the spectrum consisting of the unfolded continuum
plus the folded line were shifted to the source redshift.  We
selected bin widths of 0.25\,keV for $E\le$\,8\,keV and 2.5\,keV for
$E\ge$\,8\,keV in this process, and used a Monte Carlo method to
redistribute the observed events into the new spectral bins.  The 43
spectra with the folded line are then co-added.  The unfolded
continuum model spectra were also co-added.  Finally, the ratio of
the two spectra is calculated, as 
shown in Figure\,\ref{fig: 43pg ratio}. A prominent narrow core at
$\sim$6.4\,keV and a very weak low-energy wing are seen.

\subsection{Simultaneous Fit to All Spectra} \label{subsec: fitting 43pg}

\begin{figure}[htbp]
\begin{center}
\includegraphics[angle=-90,scale=.30]{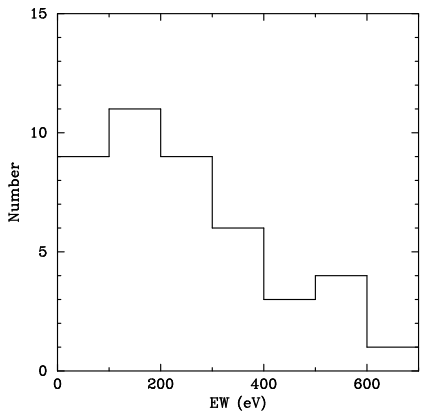}
\caption{Distribution of equivalent widths of Fe K line in 43 PG
quasars. A single Gaussian model is assumed.
}
\label{fig: EW distribution all}
\end{center}
\end{figure}

In contrast to previous studies, we employ simultaneous fits rather than fits 
to stacked spectra.  Since the Fe K line in each quasar has a different peak 
energy in the observed frame according to their redshifts, a technique that 
properly treats the energy-dependent detector response is necessary.
We fitted all spectra simultaneously in order to determine mean Fe K
line parameters. The spectral parameters of the best-fit continuum
models for each object were used as initial parameters in the spectral
fits. All the continuum parameters were left free, while the
parameters for the edge model are fixed at the best-fit values for
individual fits.  A Gaussian model with common peak energy and width
is added to the continuum model. The normalizations of the Gaussian
component were left free individually.  The fitting results are shown in
Table\,\ref{tab: fitting 43pg}. The best-fit peak energy and the
dispersion of the Gaussian line are $E=6.48^{+0.05}_{-0.04}$\,keV and
$\sigma=0.36^{+0.08}_{-0.08}$\,keV, respectively. The mean of the distribution 
of the equivalent widths, shown in Figure\,\ref{fig: EW distribution all}, is 
248\,eV, with a standard deviation of 168\,eV.

A relativistic line model ({\tt diskline} model in XSPEC) instead of a
Gaussian was also examined. We fixed several parameters; the peak
energy at 6.4\,keV, the outer radius $R_{\rm out}$ at 500\,$R_{\rm
s}$, where $R_{\rm s}$ is the Schwarzschild radius, the emissivity
index $\beta$ at $-$2, and the inclination angle of the disk $i$ at 30
degrees. The free parameters are the inner radius $R_{\rm in}$ and the
intensity of the line. Again, the parameters in the continuum model
were initially set to the best-fit values in the individual fits and
left free except for the edge component. The results are summarized in
Table\,\ref{tab: fitting 43pg}. The best-fit inner radius $R_{\rm
in}$ is $3.0^{+3.3}_{-0.0}\,R_{\rm s}$. The lower boundary is fixed
at $3R_{\rm s}$, which is the minimum value allowed in the {\tt
diskline} model assuming a Schwarzschild black hole.  The mean and
standard deviation of the equivalent widths are 210\,eV and 133\,eV,
respectively.

\begin{figure}[htbp]
\begin{center}

\includegraphics[angle=-90,scale=.30]{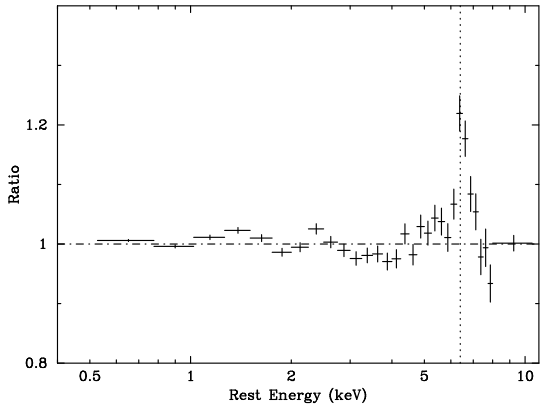}
\caption{Data-to-model ratio for co-added spectrum 
of 43 PG quasars. Vertical dotted line is at 
6.4\,keV.} 
\label{fig: 43pg ratio}
\end{center}
\end{figure}

The line profile in Figure\,\ref{fig: 43pg ratio} shows a prominent
narrow core and a very weak low-energy wing. Such a profile may be 
indicative of the presence of multiple line components, and a single
Gaussian or a diskline profile may not be a good representation. Therefore, we
examined a two-component line model consisting of either two Gaussians or a
combination of a Gaussian and a diskline component.  With regards to the 
double-Gaussian model, although the $\chi^2$ improved by 103 for 3 
additional parameters, not all the line parameters were tightly
constrained. Thus, we first fixed the continuum component to obtain
a rough value of the Gaussian widths, and then the widths were fixed at
these values to constrain the equivalent widths of the lines. All the
parameters except for the widths were left free in the latter
fit. Note that the normalizations of the lines were determined for
each object.  The resulting peak energies of the broad and narrow
lines are 5.06\,keV and 6.52\,keV, respectively (see Table\,\ref{tab:
fitting 43pg}). The mean equivalent width for the narrow component is 261 eV,
and for the broad component it is 150 eV, although the standard
deviations of the distributions are large (171 and 181 eV,
respectively). The distributions suggest that the narrow component is
commonly seen and that the broad line is present in a fraction of the
sample.

The Gaussian plus diskline model was then examined. We fixed the peak
energy and the dispersion of the Gaussian to the best-fit parameters of the
narrow component in the double-Gaussian fit. The fixed parameters for the 
diskline component were the same as those in the single diskline fit presented
above. The free parameters are the inner radius and the normalization
of the diskline, and the parameters for the Gaussian and
continuum. As summarized in Table\,\ref{tab: fitting 43pg}, the best-fit
inner radius is $R_{\rm in}$  = $3.0^{+5.9}_{-0.0}\,R_{\rm s}$, and the
mean equivalent width of the diskline is EW=102\,eV. The large standard
deviation (181 eV) again suggests that the diskline is required in only a fraction of the sample.

\begin{table*}[htbp]
\begin{center}
\caption{Results of Spectral Fit.\label{tab: fitting 43pg}}

\begin{tabular}{ccccccccccc}
\tableline\tableline
   & \multicolumn{3}{c}{Gaussian} & \multicolumn{6}{c}{Diskline} \\ \cline{5-10} 
 Model & Energy & $\sigma$ & EW\tablenotemark{a} &
 Energy &$R_{\rm in}$ & $R_{\rm out}$ & $\beta$ & $i$ & EW\tablenotemark{a} & $\chi^2$/d.o.f. \\
       & (keV) & (keV) & (eV) & (keV) & ($R_{\rm s}$) & ($R_{\rm s}$) & & (deg) &(eV) &\\
\tableline
 Gaussian & $6.48^{+0.05}_{-0.04}$ & $0.36^{+0.08}_{-0.08}$ & 
 $248 (168)$   & $-$ & $-$ & $-$ & $-$ & $-$ &$-$ & 8958/8129\\
 Diskline & $-$ & $-$ & $-$ & 6.4f &
 $3.0^{+3.3}_{-0.0p}$ & 500f & $-$2f & 30f & $210(133)$ & 8975/8130 \\
 Gaussian$+$Gaussian & $5.06^{+0.17}_{-0.13}$ & 0.61f & 150(181) &
 $-$ &$-$ &$-$ & $-$ & $-$ & $-$ &   \\
                     & $6.52^{+0.05}_{-0.04}$ & 0.34f & 261(171) &
 $-$ &$-$ &$-$ & $-$ & $-$ & $-$ & 8853/8086 \\
 Gaussian$+$Diskline & 6.52f & 0.34f & 155(181) &
 6.4f & $3.0^{+2.9}_{-0.0p}$ & 500f & $-$2f & 30f & $102 (122)$ & 8912/8087 \\
\tableline

\end{tabular}

\tablenotetext{a}{Numbers in parentheses are standard deviation 
of distribution of best-fit values.}
\tablenotetext{b}{``p'' denotes pegged parameter.}
\tablenotetext{c}{``f'' denotes fixed parameter.}

\end{center}
\end{table*}

\subsection{Eddington Ratio Dependence of the Fe Line Profile}\label{sub: eddington dependence}

We now turn to the Eddington ratio dependence of the Fe line
profile. The 43 quasars were divided into four groups according to the
Eddington ratio as follows: (1)\,$1.00 \le L_{\rm bol}/L_{\rm Edd}$,
(2)\,$0.40 \le L_{\rm bol}/L_{\rm Edd}\le 1.00$, (3)\,$0.18 \le L_{\rm
bol}/L_{\rm Edd}\le 0.40$, and (4)\,$L_{\rm bol}/L_{\rm Edd} \le 0.18$
(see Table\,\ref{tab: the sample}). The spectra in each group were
co-added as in Section \,\ref{subsec: fitting 43pg}
for the purposes of presentation; Figure\,\ref{fig: 10pg
ratio} shows the resulting data-to-continuum model ratios.  As the 
Eddington ratio increases, the width of the Fe K line appear to becomes broader 
and its peak energy rises.

\begin{figure}[htbp]
\begin{center}
\includegraphics[angle=-90,scale=.30]{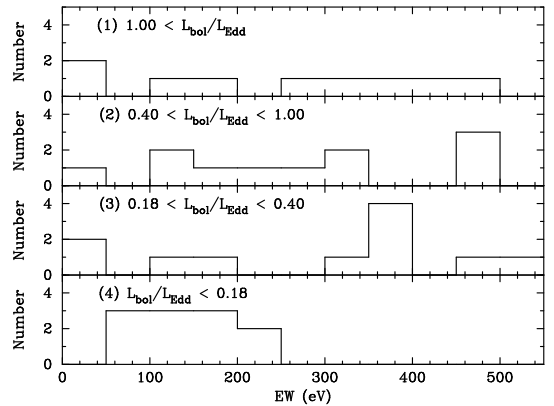}

\caption{
Distribution of Fe line equivalent width in each Eddington ratio group.}
\label{fig: EW distribution}
\end{center}
\end{figure}

Following the procedure outlined in Section \,\ref{subsec: fitting
43pg}, we fitted the spectra in each group simultaneously to characterize the
Fe K line quantitatively.
The following four models were examined for the Fe K line
component: single Gaussian, single diskline, double Gaussians, and
a Gaussian plus diskline model. The results of the spectral fits are
shown in Table\,\ref{tab: fitting 10pg}, and the Eddington ratio
dependence of the peak energy and the width of the line obtained from
the single Gaussian model fits are shown in Figure\,\ref{fig:
dependence of the param}. We again find that the width and peak energy
of the line become broader and higher as the Eddington ratio increases.
The peak energy gradually increases from 6.37 keV in group
4 to 6.77 keV in group 1.  The width is relatively narrow
($\sigma$ = 0.15\,keV) in group 4, but definitely resolved 
($\sigma_{\rm inst}$ = 0.07\,keV), and becomes wider ($\sigma$ = 0.68 keV in
group 1) as $L_{\rm bol}/L_{\rm Edd}$ increases.  The distributions of the
equivalent widths for each group obtained from the Gaussian fits
are shown in Figure\,\ref{fig: EW distribution}; the mean equivalent
widths in each group are 280, 276, 287, and 131\,eV, respectively.


A diskline model was examined next. We fixed several diskline
parameters.  The line center energy was chosen to be 6.4 keV for 
groups 3 and 4 and 6.7 keV for groups 1 and 2, for which
the Gaussian fits suggest that the line is from ionized Fe. The outer
radius $R_{\rm out}$, the emissivity index $\beta$, and the
inclination angle $i$ were fixed at 500\,$R_{\rm S}$, $-$2, and 30
degrees, respectively. The free parameters are the inner radius
$R_{\rm in}$ and the normalization. The parameters for the continuum
model were also left free, except for those of the edge component. The
results are shown in Table\,\ref{tab: fitting 10pg}. The resulting
values of $\chi^2$ are similar to or slightly worse than those for the
Gaussian fits. The best-fit inner radius is $R_{\rm in}\approx 3R_{\rm
S}$, which is the last stable orbit of a Schwartzschild black hole.

We also examined a two-component line model since a strong narrow 
core and  a broad wing are visible in groups 3 and 4, and possibly
in group 2.  A double-Gaussian model
was first tested in the same way as in Section\,\ref{subsec: fitting
43pg}; the results are shown in Table\,\ref{tab: fitting 10pg}. 
The resulting $\chi^2$ values are similar to or only slightly better
than those of the single-Gaussian fit, indicating that the additional
broad line is statistically not required. The distributions
of the EW are broad and indicate that some objects may have a broad
component.

A Gaussian plus diskline model was then used in order to constrain the
equivalent width of the diskline, in which the inner radius is fixed
at 3$R_{\rm S}$. The peak energy and the dispersion of the Gaussian were
fixed at the best-fit parameters for the narrow component in the
double-Gaussian fit. The inner radius of the disk was also fixed at
3$R_{\rm S}$, and the other parameters in the diskline model were fixed to
the same values as in the single-diskline fits described above. Only the
normalizations of the Gaussian and diskline were left free. The
parameters for the continuum model were treated in a similar way as
in Section 3.3. The results are summarized in Table\,\ref{tab:
fitting 10pg}. The resulting values of $\chi^2$ are similar to or only
slightly better than those for the single-Gaussian or diskline fits. The
mean equivalent widths of the diskline in each group are 99,
176, 132, and 67\,eV, respectively, although the standard deviations as
large. The large standard deviations again suggest that the broad diskline
is not ubiquitous. 

\begin{figure}[htbp]
\begin{center}
\includegraphics[angle=-90,scale=1.0]{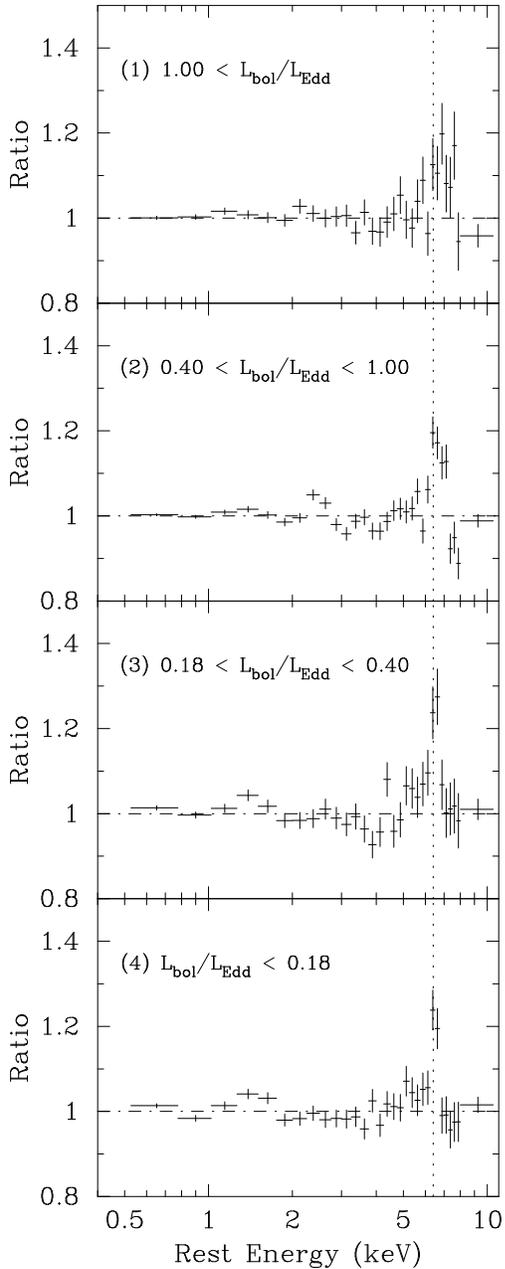}
\caption{Eddington ratio dependence of the Fe line profile for each of the 
four groups.  The vertical dotted line is at 6.4\,keV.}
\label{fig: 10pg ratio}
\end{center}
\end{figure}


\begin{table*}[htbp]
\begin{center}
\caption{Results of Spectral Fit in Each Eddington Ratio Group. \label{tab: fitting 10pg}}
\begin{tabular}{cccccccccccc}
\tableline\tableline
 &  & \multicolumn{3}{c}{Gaussian} & \multicolumn{6}{c}{Diskline} \\ \cline{6-11} 
Group & Model & Energy & $\sigma$ & EW\tablenotemark{a} &
 Energy &$R_{\rm in}$ & $R_{\rm out}$ & $\beta$ & $i$ & EW\tablenotemark{a} & $\chi^2$/d.o.f. \\
 &       & (keV) & (keV) & (eV) & (keV) & ($R_{\rm s}$) & ($R_{\rm s}$) & & (deg) &(eV) &\\
\tableline
1 & Gaussian & $6.77^{+0.30}_{-0.30}$ & $0.68^{+0.38}_{-0.24}$ & 
 $280 (199)$   & $-$ & $-$ & $-$ & $-$ & $-$ &$-$ & 2308/2237\\
 & Diskline & $-$ & $-$ & $-$ & 6.7f &
 $3.0^{+20.0}_{-0.0p}$ & 500f & $-$2f & 30f & $202 (176)$ & 2316/2238 \\
& Gaussian$+$Gaussian & $6.58^{+0.14}_{-0.56}$ & 0.42f & $119 (152)$ &
 $-$ &$-$ &$-$ & $-$ & $-$ & $-$ &   \\
&                     & $7.47^{+0.58}_{-1.44}$ & 0.48f & $257 (395)$ &
 $-$ &$-$ &$-$ & $-$ & $-$ & $-$ & 2305/2248 \\
 & Gaussian$+$Diskline & 7.47f & 0.48f & $208 (281)$ &
 6.7f & 3.0f & 500f & $-$2f & 30f & $99 (158)$ & 2304/2250 \\
\tableline
2 & Gaussian & $6.54^{+0.08}_{-0.08}$ & $0.35^{+0.14}_{-0.07}$ & 
 $276 (160)$   & $-$ & $-$ & $-$ & $-$ & $-$ &$-$ & 2636/2197\\
 & Diskline & $-$ & $-$ & $-$ & 6.7f &
 $3.0^{+1.9}_{-0.0p}$ & 500f & $-$2f & 30f & $285 (171)$ & 2662/2198 \\
& Gaussian$+$Gaussian & $5.02^{+0.30}_{-0.25}$ & 0.38f & $90 (137)$ &
 $-$ &$-$ &$-$ & $-$ & $-$ & $-$ &   \\
&                     & $6.52^{+0.07}_{-0.07}$ & 0.24f & $183 (96)$ &
 $-$ &$-$ &$-$ & $-$ & $-$ & $-$ & 2634/2218 \\
 & Gaussian$+$Diskline & 6.52f & 0.24f & $68 (80)$ &
 6.7f & 3.0f & 500f & $-$2f & 30f & $176 (203)$ & 2646/2220 \\
\tableline
3 & Gaussian & $6.45^{+0.09}_{-0.10}$ & $0.37^{+0.25}_{-0.12}$ & 
 $287 (179)$   & $-$ & $-$ & $-$ & $-$ & $-$ &$-$ &1796/1774 \\
 & Diskline & $-$ & $-$ & $-$ & 6.4f &
 $3.3^{+8.0}_{-0.3p}$ & 500f & $-$2f & 30f & $259 (149)$ & 1801/1775 \\
& Gaussian$+$Gaussian & $5.55^{+0.46}_{-0.87}$ & 0.48f & $96 (123)$ &
 $-$ &$-$ &$-$ & $-$ & $-$ & $-$ &   \\
&                     & $6.49^{+0.09}_{-0.08}$ & 0.25f & $173 (124)$ &
 $-$ &$-$ &$-$ & $-$ & $-$ & $-$ & 1800/1798 \\
 & Gaussian$+$Diskline & 6.49f & 0.25f & $83 (114)$ &
 6.4f & 3.0f & 500f & $-$2f & 30f & 132 (132) & 1803/1800 \\
\tableline
4 & Gaussian & $6.37^{+0.05}_{-0.04}$ & $0.15^{+0.06}_{-0.02}$ & 
 $131 (59)$   & $-$ & $-$ & $-$ & $-$ & $-$ &$-$ &2184/1935 \\
 & Diskline & $-$ & $-$ & $-$ & 6.4f &
 $3.0^{+24.2}_{-0.0p}$ & 500f & $-$2f & 30f & $179 (68)$ & 2186/1936 \\
& Gaussian$+$Gaussian & $5.90^{+0.37}_{-0.56}$ & 0.70f & $104 (108)$ &
 $-$ &$-$ &$-$ & $-$ & $-$ & $-$ &   \\
&                     & $6.37^{+0.04}_{-0.03}$ & 0.09f & 82 (70) &
 $-$ &$-$ &$-$ & $-$ & $-$ & $-$ & 2173/1957 \\
 & Gaussian$+$Diskline & 6.37f & 0.09f & $75 (76)$ &
 6.4f & 3.0f & 500f & $-$2f & 30f & $67 (73)$ & 2180/1959 \\
\tableline

\end{tabular}
\tablenotetext{a}{Numbers in parentheses are standard deviation 
of distribution of best-fit values.}
\tablenotetext{b}{``p'' denotes pegged parameter.}
\tablenotetext{c}{``f'' denotes fixed parameter.}
\end{center}
\end{table*}

\begin{figure}[htbp]
\begin{center}
\includegraphics[angle=-90,scale=.30]{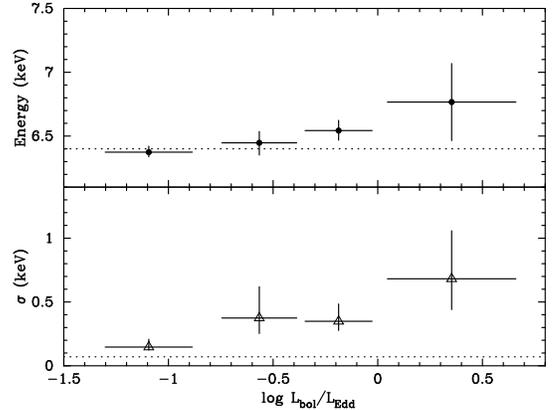}
\caption{Eddington ratio dependence of the peak energy and the width of
 the line. The horizontal dotted line on the top panel is at 6.4\,keV,
and that on the bottom panel is at 0.07\,keV, the instrumental
resolution of {\it XMM-Newton}.}

\label{fig: dependence of the param}
\end{center}
\end{figure}

\section{Discussion}

\subsection{Origin of the Eddington Ratio Dependence} \label{sub:eddington dependence}

We examined mean Fe K line shapes for the four groups sorted by the
Eddington ratios, and found that the peak energy, width, and
equivalent width become higher (6.4 to 6.8\,keV), broader (0.1 to
0.7\,keV), and larger (130 to 280\,eV), respectively, as the Eddington
ratio increases. In this section, we discuss possible reasons of
the observed Eddington ratio dependence of the Fe line profile.



The line center energy of the Fe K line is governed by a combination of the 
Doppler effect caused by the orbital motion, gravitational redshift, and 
the ionization state of the emitter. An Fe line from a rotating disk around a 
black hole
shows a double-peaked profile with a peak energy that depends on the
inclination angle of the disk. Since each of the four groups we used
contains from 10 to 11 objects, it is unlikely that any one of the groups
has a mean inclination angle considerably different from those of other
groups. Therefore, the different line center energies obtained for 
the four groups probably reflect different ionization
states rather than different mean disk inclination angles.
The Eddington ratio dependence of the peak energy thus strongly
suggests that the ionization state of the accretion disk depends on
the Eddington ratio. Many aspects of an ionized disk have been
studied theoretically over the last decade. Calculations have been performed for
constant-density atmospheres (e.g., Matt et al. 1993; Ross \& Fabian 1993;
$\dot{\rm Z}$ycki et al. 1994), as well as for atmospheres in 
hydrostatic equilibrium (Nayakshin et
al. 2000; Ballantyne et al. 2001). These authors calculated
reflection spectra from an ionized disk for different conditions.
In the lowest ionization regime, the most prominent feature is the
cold iron line at 6.4\,keV. As the mass accretion rate goes up, the
ionized stage becomes higher, and the Fe K line is dominated by 
Fe XXV and XXVI, and K lines from the lighter elements emerge in the
$0.3-3$\,keV band. At very high accretion rates, the surface of the disk
is highly ionized so that the only noticeable line is a Compton-broadened 
Fe K line peaking at $\sim$7\,keV. The Eddington ratio
dependence of the peak energy and the width of the Fe K line we
observed in the PG quasar sample is in good agreement with the
predictions of ionized disk models.


The large EW in the high-Eddington ratio groups is also likely to
be  related to the
ionization state of the accretion disk. 
Calculations of the reflected X-rays from an ionized disk have shown
that the EW of the Fe K line can be higher than in the case of a disk in a
low-ionization state.  The resonant trapping effect plays a role in
the intermediate-ionization state higher than
\ion{Fe}{18}. The EW increases again from \ion{Fe}{20}, reaching a 
maximum value of $250-300$\,eV (Matt
et al.  1993; Ross \& Fabian 1993; $\dot{\rm Z}$ycki \& Czerny 1994). The observed line center energies
and EWs for the three groups with a largest Eddington ratios are
consistent with the latter expectation, although the scatter of the actual 
EW distributions is large. The reduction of EW caused by resonant trapping 
is not
clearly seen in our data. This may be due to the coarse binning of 
the Eddington ratio we have adopted. Future analysis of a larger sample may be
able to see the ionized disk in an intermediate-ionization state.


The Eddington ratio dependence of the line width also suggests a change of
the geometrical structure of the disk, if the velocity width indeed reflects 
emission from the inner part of an accretion disk broadened by Doppler effects 
and gravitational redshift. Although the line widths obtained from the
single-Gaussian fits are consistent with systematic variations in the 
inner disk radius $R_{\rm in}$, such a trend is not clear in the fits 
obtained using the diskline model.  It is possible that the broader line 
widths at high Eddington ratios may in part be accounted for by a blend of
lines from Fe in different ionization states.

\subsection{Comparison with Average Fe Line Profile of AGNs in Lockman Hole}

Streblyanska et al. (2005) derived an average Fe line profile of 53
type 1 AGN detected in a 770 ks observation of the Lockman Hole with
{\it XMM-Newton}. Their average spectrum shows a broad line-like
feature ($\sigma=0.69$\,keV) peaking at 6.4\,keV with an equivalent
width of EW = 420\,eV. The profile is asymmetric, with a significant wing
seen toward low energies.  The line width ($\sigma$ = 0.36\,keV) 
we obtained for the PG quasar sample is significantly narrower, and the 
equivalent width (EW$\approx$250\,eV) smaller, than that of
Streblyanska et al.'s profile.  We discuss possible
causes for the difference.

\subsubsection{Contribution of Absorbed Quasars}

Appropriate modeling of the continuum is essential to derive a correct
Fe line profile. A single power-law model was assumed in the analysis
of the AGNs in the Lockman Hole. Since the spectra of many AGNs show a
signature of absorption by ionized and/or cold matter, a simple power
law may not be a good approximation of the underlying
continuum. Indeed, four quasars in our sample are absorbed by a large
column density ($N_{\rm H}\approx 10^{23}$ cm$^{-2}$).  If there are some
highly absorbed objects in the Lockman Hole sample, using a single
power-law model for the continuum may mimic a broad-line profile.

The flux of a highly absorbed power law with a column density of
$10^{23}\,{\rm cm}^{-2}$ at 5\,keV is about 50\% of that of an
unabsorbed power law. If the fraction of highly absorbed quasars
is 1/10, their contribution to the low-energy side of the Fe K
line profile around 5 keV is less than 5\%.  The contribution of the absorbed
continuum depends on the fraction of absorbed quasars. Mateos et
al. (2005) measured the absorbed fraction for their sample of Lockman
Hole AGNs with sufficient X-ray counts. They found that seven out of
46 type 1 AGNs show a highly absorbed spectrum. This fraction is
similar to that in our PG quasar sample. Note, however, that
Streblyanska et al.'s sample may contain more highly absorbed objects
because their flux limit is lower than that used by Mateos et al.
(2005) and X-ray surveys have shown that objects with lower flux
tend to have a harder spectrum. If the absorbed fraction is 10\% or higher,
rather than 5\%, in the Lockman hole sample, a considerable fraction
of the red wing might be explained by the contribution of highly
absorbed emission.

\subsubsection{Distribution of Luminosities}

\begin{figure}[htbp]
\begin{center}
\includegraphics[angle=-90,scale=0.3]{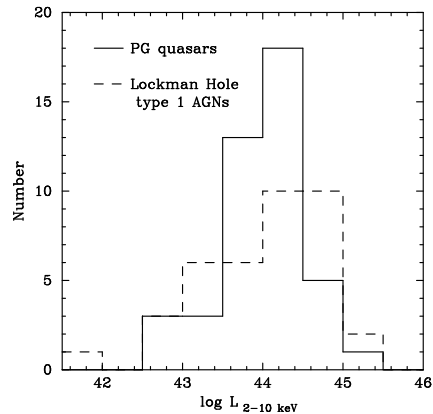}
\caption{Distribution of luminosities for the PG quasars in our sample and for 
the AGNs in the Lockman Hole. Mean luminosities are $L_{\rm
 PG}$=$10^{43.98}$\,ergs\,s$^{-1}$ and $L_{\rm LH}$= $10^{44.00}$\,ergs\,s$^{-1}$, 
respectively.}
\label{fig: lumin distri}
\end{center}
\end{figure}

The physical state of an accretion disk may depend on luminosity.  If
the mean luminosity of our sample is fairly different from that of the
Lockman Hole AGNs, it may introduce a difference in the Fe K line
profile. In order to examine this possibility, we compared the
distributions of the $2-10$\,keV luminosity between the two samples
(Fig.\,\ref{fig: lumin distri}).  The luminosities of the Lockman Hole
AGNs were taken from Mainieri et al. (2002). They
analyzed 39 type 1 AGNs in the {\it XMM-Newton}\ observation of the Lockman Hole
field. Most of the 53 type 1 AGNs in Streblyanska et al. (2005) have
been analyzed by Mainieri et al. (2002).  The two distributions
are very similar, and their mean luminosities are essentially identical: the
mean luminosity of our sample is $10^{43.98}$\,ergs\,s$^{-1}$, while
that of the Lockman Hole AGNs is $10^{44.00}$\,ergs\,s$^{-1}$. We conclude
that luminosity differences is not the main cause of the
difference of the line profiles.

\subsubsection{Artificial Distortion of Spectra}

Yaqoob (2005) pointed out a spurious effect in adding spectra of
faint sources. He showed that conventional methods for averaging low
signal-to-noise ratio X-ray spectra result in a weakening of emission lines,
a broad dip in the continuum above an emission line, and a spectral
hardening at the highest energies, if sources at different redshifts
are stacked. Yaqoob (2005) simulated 200 spectra from sources with
redshifts 0.5--2.5 and 2--10 keV fluxes 6--60$\times10^{-14}$ ergs
cm$^{-2}$ s$^{-1}$ and found an artifact wing at energies below the peak of 
the Fe K line. This feature occurs at the level of about 5\%--10\% at 4--6 keV.  This
kind of spectral distortion may affect the Fe line shape in the analysis of
Streblyanska et al. (2005) since their sample has a relatively low
average flux ($F_{\rm 2-10\,keV}\approx 10^{-14.0}\,{\rm
ergs~s^{-1}~cm^{-2}}$), compared to our PG quasar sample ($F_{\rm
2-10\,keV}\approx 10^{-11.5}\,{\rm ergs~s^{-1}~cm^{-2}}$).

\section{Conclusions}

 We presented the average Fe K line profile of 43 PG quasars observed
by {\it XMM-Newton}. Our findings are summarized as follows.

\begin{itemize}

 \item 

The spectra of 39 objects are well reproduced by a canonical 
model consisting of
a power law, a blackbody, an edge in the case of objects with 
evidence for warm absorption,
and a Gaussian, all modified by Galactic absorption,
while that of the remaining four objects require additional absorption
with a column density of $10^{23} {\rm cm}^{-2}$ to the power-law
component.

 \item An average spectrum of the 43 quasars shows a prominent narrow
 Fe K line ($E$ = 6.48\,keV, $\sigma$ = 0.36\,keV, and EW = 248\,eV).

 \item The average Fe K line is well represented by double Gaussians
 or a combination of a Gaussian and a diskline model. The best-fit
 inner radius of the disk is 3$R_{\rm S} \,(\le 5.9R_{\rm S})$ and the
 mean EW is
$\sim$100\,eV. The large standard deviation of the EW distribution 
indicates that a broad disklike line is not ubiquitous.

 \item We found an Eddington ratio dependence of the Fe K line
 profile. As the Eddington ratio increases, the peak energy becomes
 higher (6.4 to 6.8\,keV), the width becomes broader (0.1 to
 0.7\,keV), and the EW is larger (130 to 280\,eV).  The higher energy
 peak and larger EW at high Eddington ratio can be explained by the
 ionization of the accretion disk.  The systematic broadening of the line 
with increasing Eddington ratio may indicate a systematic decrease of the 
inner radius of the disk, or perhaps blending from a combination of Fe lines
 in different ionization states.

 \item The line width we obtained is narrower and the equivalent width smaller
 than those derived from the stacked spectrum
 of AGNs in the Lockman Hole ($\sigma$ = 0.69\,keV, EW = 420\,eV).  We attribute 
these differences to either an increased contribution of absorbed sources 
in the Lockman Hole composite spectrum or to artifacts introduced in 
averaging faint sources.

\end{itemize}

\appendix

\section{Notes on Individual Objects}\label{appe: pg qso}
\subsection{Highly Absorbed Quasars}

\begin{figure*}[htbp]
\begin{center}
\includegraphics[angle=-90,scale=0.28]{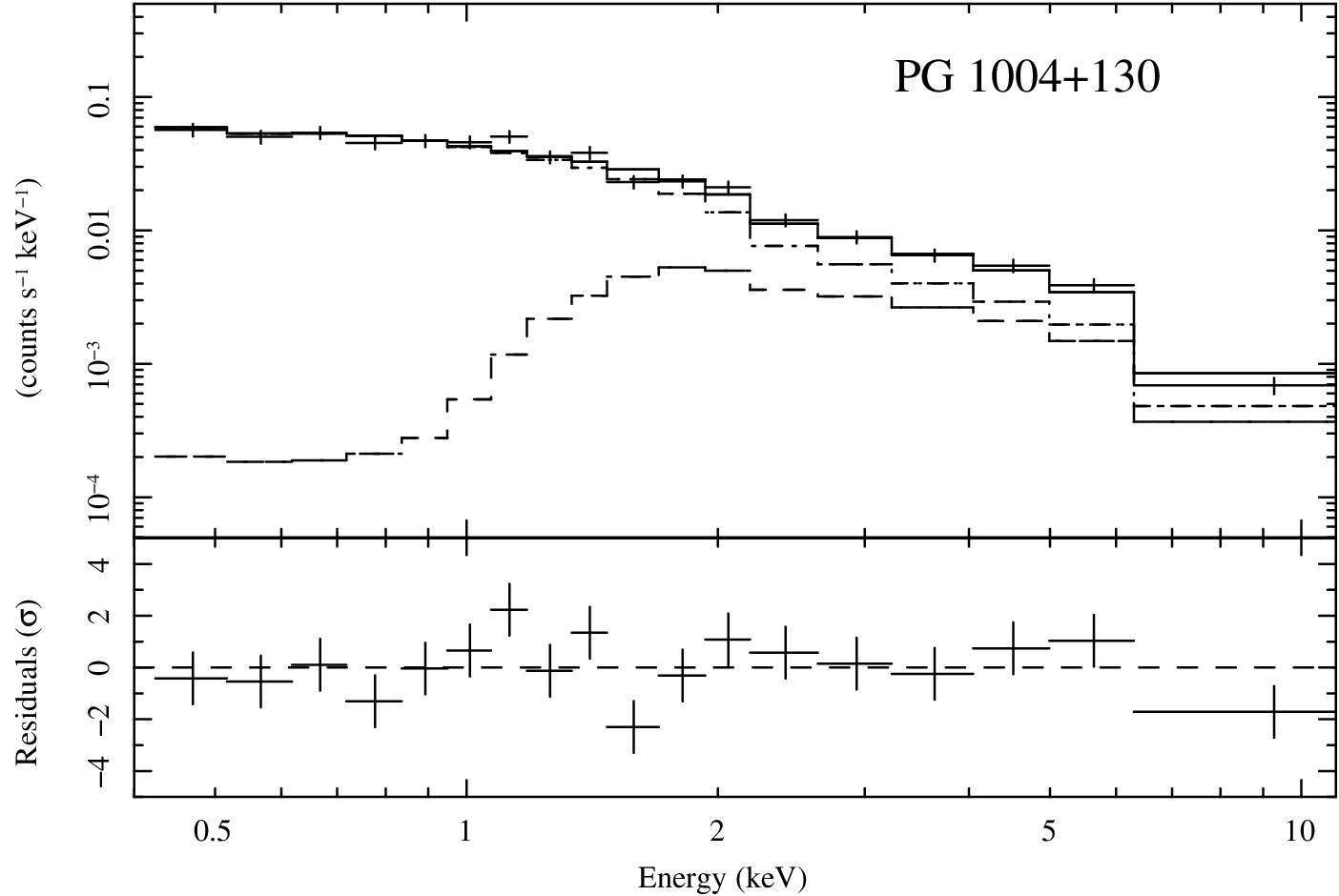}
\includegraphics[angle=-90,scale=0.28]{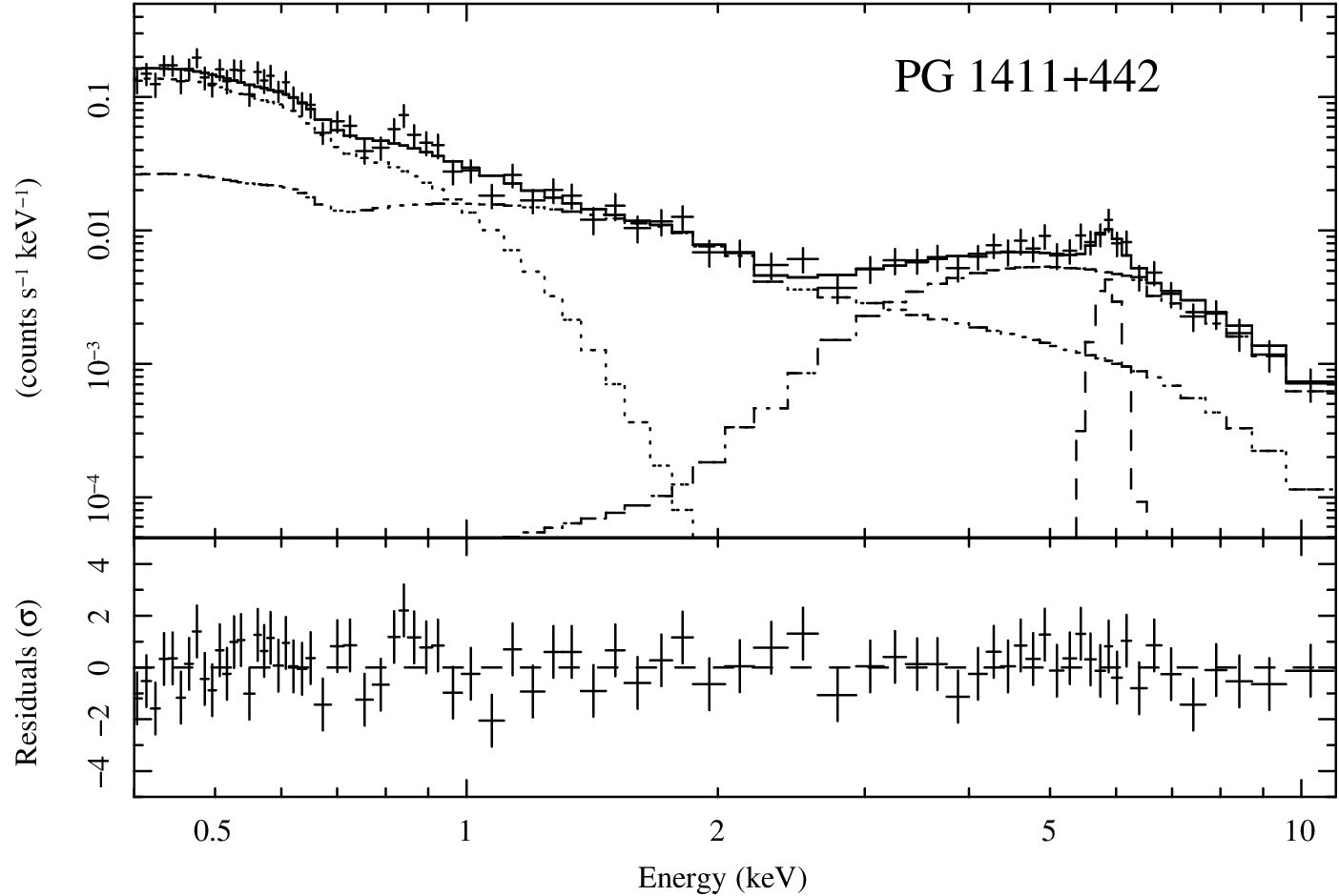}
\includegraphics[angle=-90,scale=0.28]{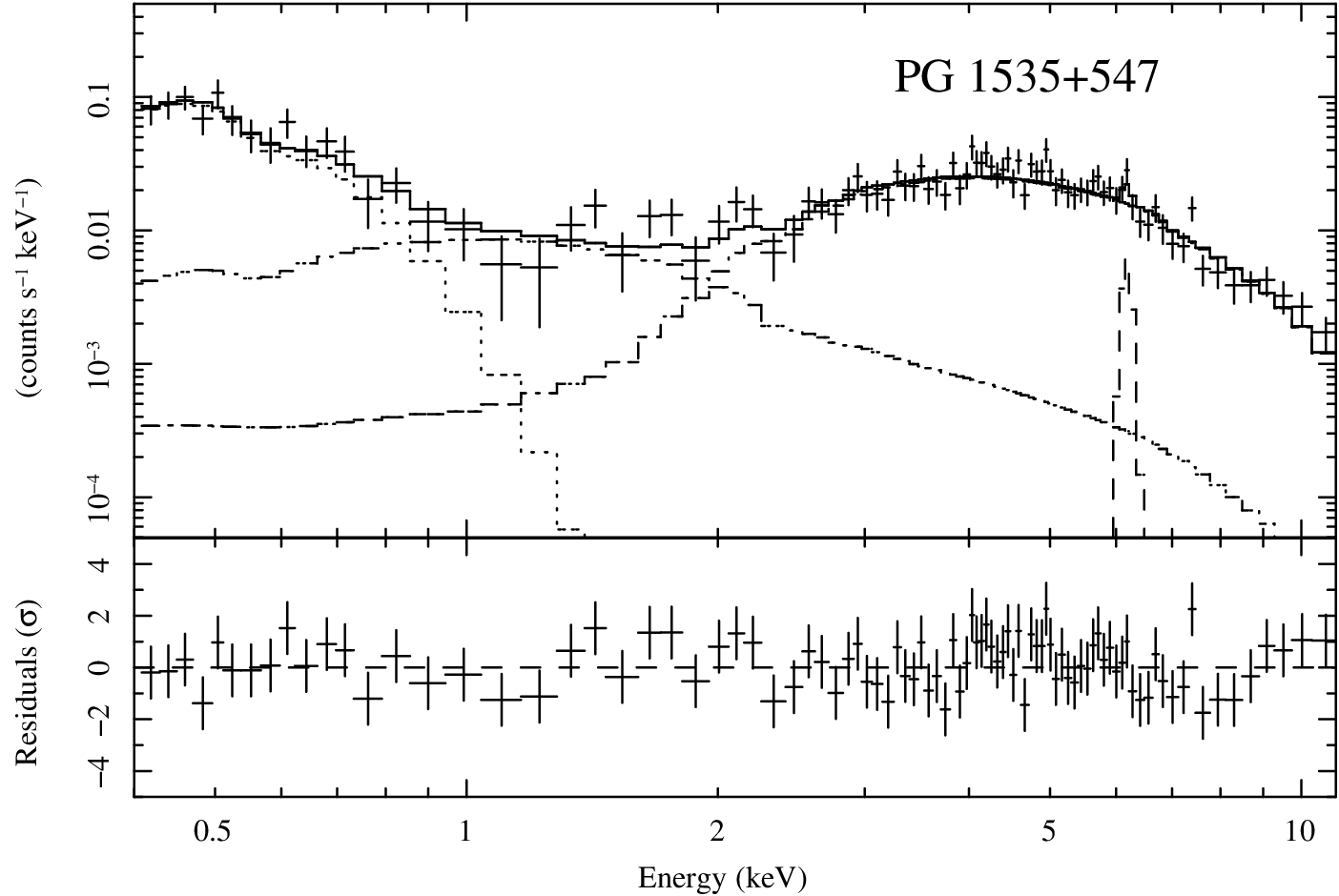}
\includegraphics[angle=-90,scale=0.28]{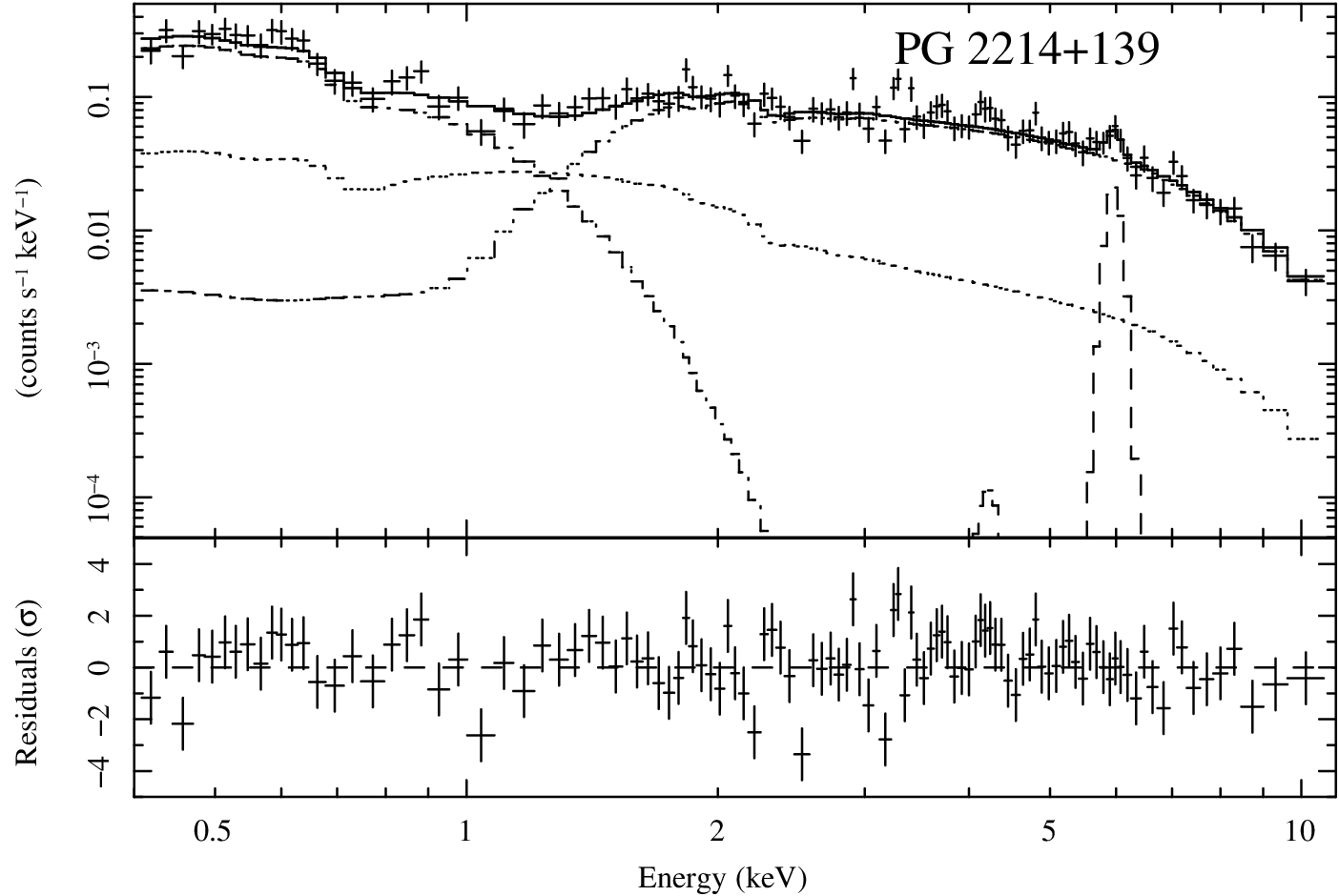}
\caption{Spectra of highly absorbed quasars.}
\label{fig: highly absorbed quasars}
\end{center}
\end{figure*}

We found four highly absorbed quasars in the 43 PG quasars. We show
their spectra in Figure\,\ref{fig: highly absorbed quasars}. They are
not well reproduced by the canonical model, and an additional highly
absorbed component is required. Therefore, we added an absorbed
power-law component. The fitting results are summarized in
Table\,\ref{tab: abs pow}. The average column densities of the
additional absorption are $\sim10^{23}$ cm$^{-2}$. Spectra of PG
1411+442 and PG 2214+139 are presented in Brinkmann et al. (2004). The
results of PG 1411+442 are consistent with ours. The reduced $\chi^2$
value is not good for PG 2214+139 ($\chi^2_\nu \approx 1.38$) because of
non-systemtaic deviations of the data from the model seen over the energy range of 0.5--4 keV. Miller et al. (2006) and
Schartel et al. (2005) found absorbed emission in PG 1004+130 and PG
1535+547, respectively. The results obtained by Miller et al. (2006)
are also consistent with ours.  Schartel et al. (2005) found the peak
energy of Fe K line at $6.00^{+0.14}_{-0.21}$\,keV and the width of
$0.26^{+0.21}_{-0.11}$\,keV, while the line center energy is lower
than 5\,keV in our analysis. Therefore, we fixed it at 6.4\,keV and
repeated the fits.  The best-fit line width is $\sigma=0.00\,(\le
0.24)$\,keV (see Table\,\ref{tab: abs pow}). 

A population of soft X-ray-weak (SXW) quasars, where the soft
X-ray flux is $\sim 10-30$ times smaller than in typical quasars (Yuan
et al. 1998), has been known. This class comprises $\sim$10\% of
optically selected quasars. Brandt et al. (2000) found 10 SXW quasars
in the Boroson \& Green (1992) sample of 87 PG quasars.  All of the highly
absorbed quasars presented here are consistent with the SXW
classification of Brandt et al. (2000). The large absorption seems to
be the cause of the X-ray weakness at least in these four objects.

\begin{deluxetable}{cccccccccc}[htbp]
\tabletypesize{\scriptsize}
\tablecaption{Spectral Parameters for Highly Absorbed Quasars. \label{tab: abs pow} }
\tablewidth{0pt}
\tablehead{
\colhead{PG } & \colhead{$\Gamma$} & \colhead{$kT$} &
 \colhead{$(N_{\rm H})_{\rm z=0}$\tablenotemark{a}} & \colhead{$N_{\rm H}$\tablenotemark{b}} &
\colhead{Edge/$\tau$} & \colhead{$E_{\rm Fe}$} & \colhead{$\sigma$} & \colhead{EW} & \colhead{$\chi^2$/d.o.f.} \\
\colhead{ } & \colhead{} & \colhead{(keV)} &
 \colhead{($10^{22}$cm$^{-2}$) } & \colhead{($10^{22}$cm$^{-2}$) } &
\colhead{(keV) } & \colhead{(keV) } & \colhead{(keV) } & \colhead{(eV) } & \colhead{}
} 
\startdata
1004+130 &1.93$^{+0.30}_{-0.36}$  &$-$  &0.06$^{+0.03}_{-0.03}$
 &5.04$^{+2.56}_{-1.69}$  &$-$ & $-$  & $-$ & $-$  &20/14   \\
  &  &  &  & &0.94$^{+0.04}_{-0.03}$/0.89$^{+0.37}_{-0.79}$ &  & &  &  \\
1411+442 &1.55$^{+0.24}_{-0.23}$  &0.13$^{+0.01}_{-0.01}$  &0.03$^{+0.05}_{-0.03}$  &16.7$^{+4.5}_{-5.6}$ &0.72$^{+0.02}_{-0.03}$/0.85$^{+0.35}_{-0.35}$ &6.35$^{+0.14}_{-0.34}$  &0.25$^{+0.60}_{-0.17}$ &481$^{+3249}_{-481}$  &63/75   \\
1535+547 &2.24$^{+0.28}_{-0.16}$  &0.07$^{+0.03}_{-0.01}$  &0.27$^{+0.14}_{-0.07}$  &11.9$^{+1.9}_{-2.1}$ &0.93$^{+0.15}_{-0.20}$/1.86$^{+1.89}_{-1.50}$ &6.4f  &0.00\,($\le$0.24) &81$^{+262}_{-81}$  &91/87   \\
2214+139 &1.49$^{+0.15}_{-0.14}$  &0.18$^{+0.02}_{-0.07}$  &0.02$^{+0.12}_{-0.02}$  &2.4$^{+1.24}_{-0.21}$  &0.73$^{+0.02}_{-0.03}$/1.13$^{+0.44}_{-0.42}$ &6.4f  &0.06$^{+0.15}_{-0.06}$ &152$^{+484}_{-152}$  &146/106   \\

\enddata
\tablenotetext{a}{Foreground absorption within our Galaxy}
\tablenotetext{b}{Absorption within quasar}
\end{deluxetable}

\subsection{Quasars Newly Analyzed in This Paper}

We summarize the results of our spectral analysis of quasars newly presented in
this paper.  Their spectra are shown in Figure\,\ref{fig: other
quasars}. We fitted the spectra with the canonical model described in
Section 3.1;  results are listed in Table\,\ref{tab: other qso}. They
are well reproduced by the model except for one case (PG 1448+273). The
soft excess component in PG 1448+273 is not explained by a single
blackbody component. We added an additional blackbody
component and obtained a good fit.

\begin{deluxetable}{ccccccccc}
\tabletypesize{\scriptsize}
\tablecaption{Spectral Parameters for Quasars Newly Analyzed in
This Paper\label{tab: other qso}}
\tablewidth{0pt}
\tablehead{
\colhead{PG } & \colhead{$\Gamma$} & \colhead{$kT$} &
 \colhead{$(N_{\rm H})_{\rm z=0}$\tablenotemark{a}} & \colhead{Edge/$\tau$} & \colhead{$E_{\rm Fe}$} & \colhead{$\sigma$} & \colhead{EW} & \colhead{$\chi^2$/d.o.f.} \\
\colhead{ } & \colhead{} & \colhead{(keV)} &
 \colhead{($10^{22}$cm$^{-2}$) } &
\colhead{(keV) } & \colhead{(keV) } & \colhead{(keV) } & \colhead{(eV) } & \colhead{}
} 
\startdata
1302$-$102 &1.66$^{+0.10}_{-0.11}$  &0.20$^{+0.02}_{-0.04}$  &0.00$^{+0.06}_{-0.00}$ &$-$  &$-$  &$-$ &$-$ &64/63 \\
1448+273   &2.25$^{+0.11}_{-0.11}$  &0.12$^{+0.01}_{-0.02}$
 &0.05$^{+0.01}_{-0.01}$ &0.66$^{+0.02}_{-0.01}$/0.27$^{+0.07}_{-0.07}$
 &6.41$^{+0.54}_{-0.24}$ &0.1f & 93$^{+142}_{-93}$ &260/290 \\
           &      &0.23$^{+0.06}_{-0.14}$  &        &  &     &     & & \\
2130+099   &1.65$^{+0.03}_{-0.02}$  &0.11$^{+0.00}_{-0.00}$  &0.05$^{+0.01}_{-0.01}$ &0.72$^{+0.02}_{-0.01}$/0.29$^{+0.06}_{-0.07}$  &6.45$^{+0.07}_{-0.12}$  &0.02$^{+0.22}_{-0.02}$ &139$^{+495}_{-139}$ &182/139 \\
2233+134   &2.17$^{+0.19}_{-0.17}$  &0.19$^{+0.03}_{-0.03}$  &0.05$^{+0.03}_{-0.03}$ &0.89$^{+0.03}_{-0.04}$/0.63$^{+0.25}_{-0.23}$  &$-$  &$-$ &$-$ &117/96 \\

\enddata
\tablenotetext{a}{Foreground absorption within our Galaxy}

\tablecomments{PG 1448+273 requires two blackbody components to represent
soft excess.
}.  

\end{deluxetable}

\subsection{Quasars Reanalyzed in This Paper}

This section shows results of quasars reanalyzed in this paper, for
which the best-fit parameters are not consistent with previously published
 results. Our results are summarized in Table\,\ref{tab: reanalyzed
 qso}. Some notes on individual objects are given below.

{\it PG 0003+199}.--- Gondoin et al. (2002) give the line center of
Fe K line at
6.0$\pm$0.14\,keV and the width of 1.2$\pm$ 0.2\,keV. 
We obtained a higher center energy $6.60^{+0.13}_{-0.02}$\,keV and
a narrower width $0.35^{+0.16}_{-0.09}$\,keV.

{\it PG 0804+761}.--- Jim\'enez-Bail\'on et al. (2005), Porquet et
al. (2004), and Page et al. (2004) show the best-fit parameters of the
line center at $6.38\pm 0.06$, $6.67^{+0.31}_{-0.37}$, and
$6.62\pm 0.14$\,keV, respectively. The peak energy measured by Porquet
et al. (2004) and Page et al. (2004b) are consistent with our
result\,($6.57\pm 0.08$\,keV).

{\it PG 1244+026}.--- The best-fit line center energy is
$6.13^{+0.12}_{-0.10}$\,keV in our analysis. Jim\'enez-Bail\'on et
al. (2005) and Porquet et al. (2004) obtained the line center energy
at $6.65^{+0.07}_{-0.18}$\,keV and $6.66^{+0.09}_{-0.07}$\,keV,
respecrively.

{\it PG 1613+658}.--- A line feature is not clearly seen at
$\sim$6.4\,keV in our spectrum. No detection is consistent with
the result of Page et al. (2004a) (EW$<$85 eV), while Page et al.
(2004b) reported EW=$96\pm59$ eV.

\begin{deluxetable}{ccccccccc}
\tabletypesize{\scriptsize}
\tablecaption{Spectral Parameters for Quasars Reanalyzed in This
 Paper. \label{tab: reanalyzed qso}}
\tablewidth{0pt}
\tablehead{
\colhead{PG } & \colhead{$\Gamma$} & \colhead{$kT$} &
 \colhead{$(N_{\rm H})_{\rm z=0}$\tablenotemark{a}} & \colhead{Edge/$\tau$} & \colhead{$E_{\rm Fe}$} & \colhead{$\sigma$} & \colhead{EW} & \colhead{$\chi^2$/d.o.f.} \\
\colhead{ } & \colhead{} & \colhead{(keV)} &
 \colhead{($10^{22}$cm$^{-2}$) } &
\colhead{(keV) } & \colhead{(keV) } & \colhead{(keV) } & \colhead{(eV) } & \colhead{}
} 
\startdata
0003+199   &2.25$^{+0.01}_{-0.01}$  &0.13$^{+0.00}_{-0.00}$
 &0.03$^{+0.00}_{-0.00}$ &0.65$^{+0.01}_{-0.01}$/0.16$^{+0.02}_{-0.02}$
 &6.60$^{+0.13}_{-0.02}$  &0.35$^{+0.16}_{-0.09}$ &192$^{+134}_{-97}$
 &626/483 \\
0804+761   &2.30$^{+0.09}_{-0.09}$  & $-$  &0.02$^{+0.01}_{-0.01}$ & $-$  &6.57$^{+0.08}_{-0.08}$  &0.1f &1406$^{+2195}_{-1353}$ &141/138 \\
1244+026   &2.54$^{+0.01}_{-0.01}$  &0.15$^{+0.01}_{-0.01}$  &0.01$^{+0.01}_{-0.01}$ &$-$  &6.13$^{+0.12}_{-0.10}$  &0.1f &642$^{+1031}_{-642}$ &197/187 \\
1613+658   &1.92$^{+0.10}_{-0.10}$  &0.13$^{+0.02}_{-0.03}$  &0.04$^{+0.02}_{-0.02}$ &0.73$^{+0.16}_{-0.07}$/0.22$^{+0.19}_{-0.20}$  &$-$  &$-$ &$-$ &109/97 \\
\enddata
\tablenotetext{a}{Foreground absorption within our Galaxy}


\end{deluxetable}

\begin{figure}[htbp]
\begin{center}
\includegraphics[angle=-90,scale=0.28]{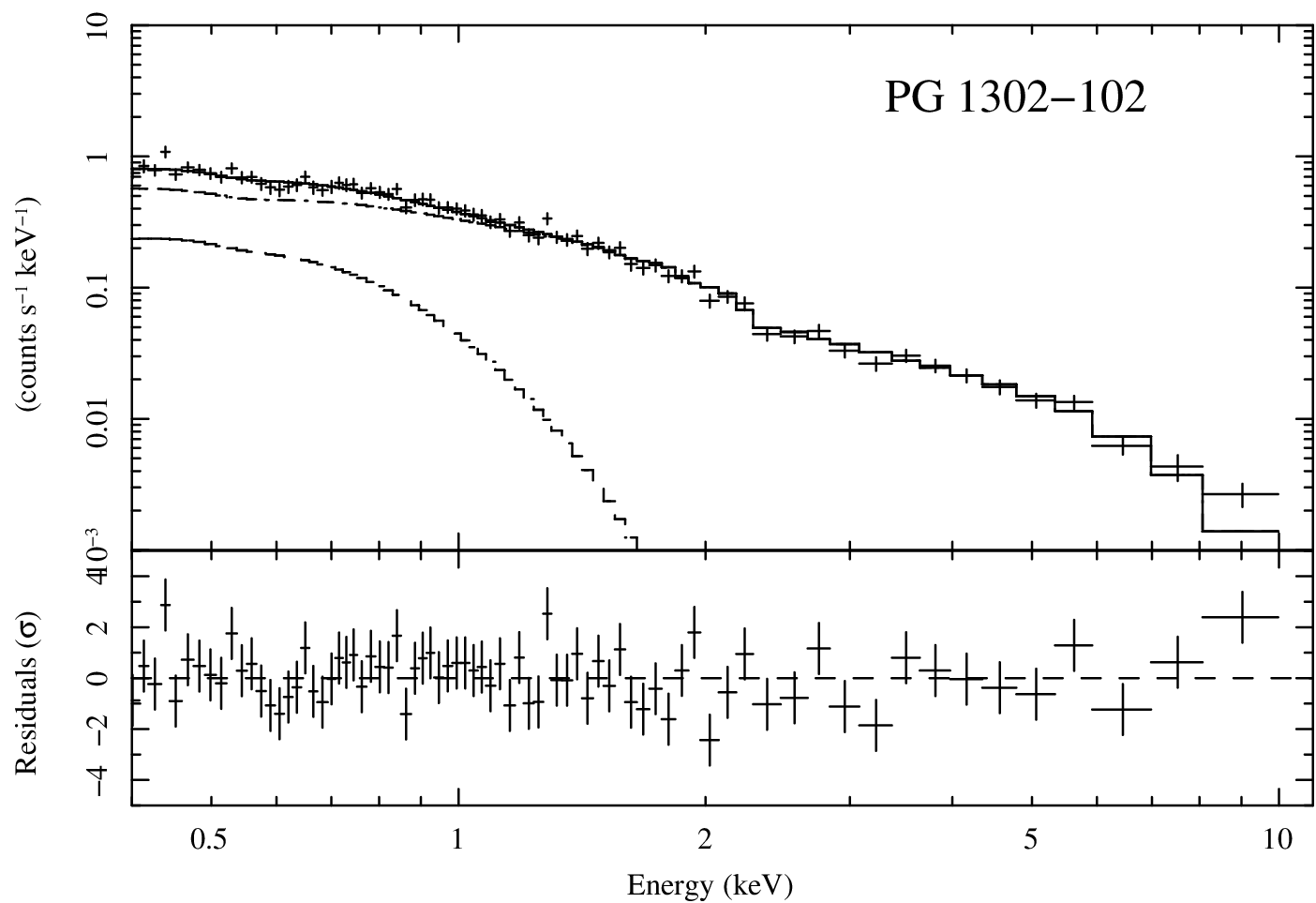}
\includegraphics[angle=-90,scale=0.28]{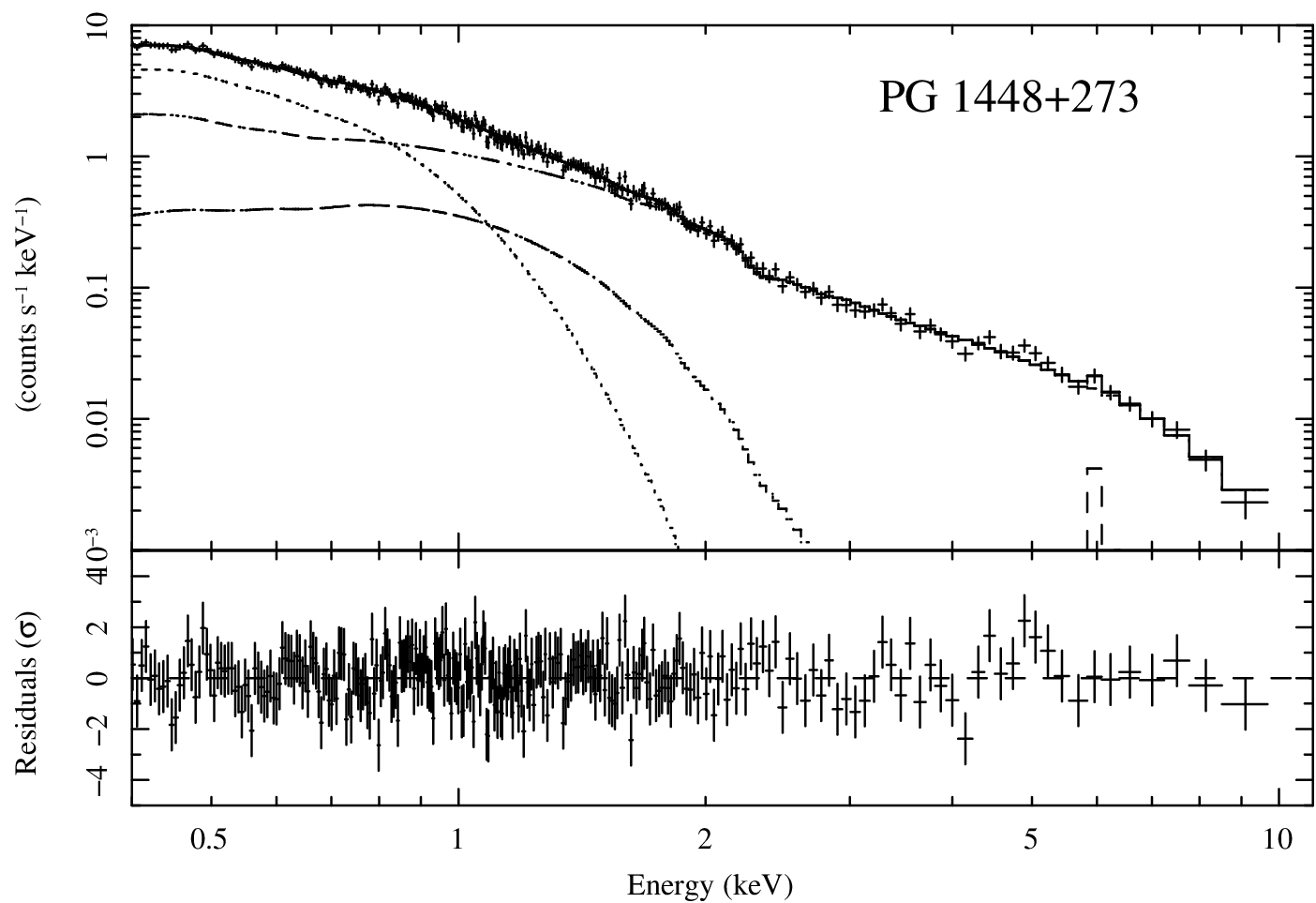}
\includegraphics[angle=-90,scale=0.28]{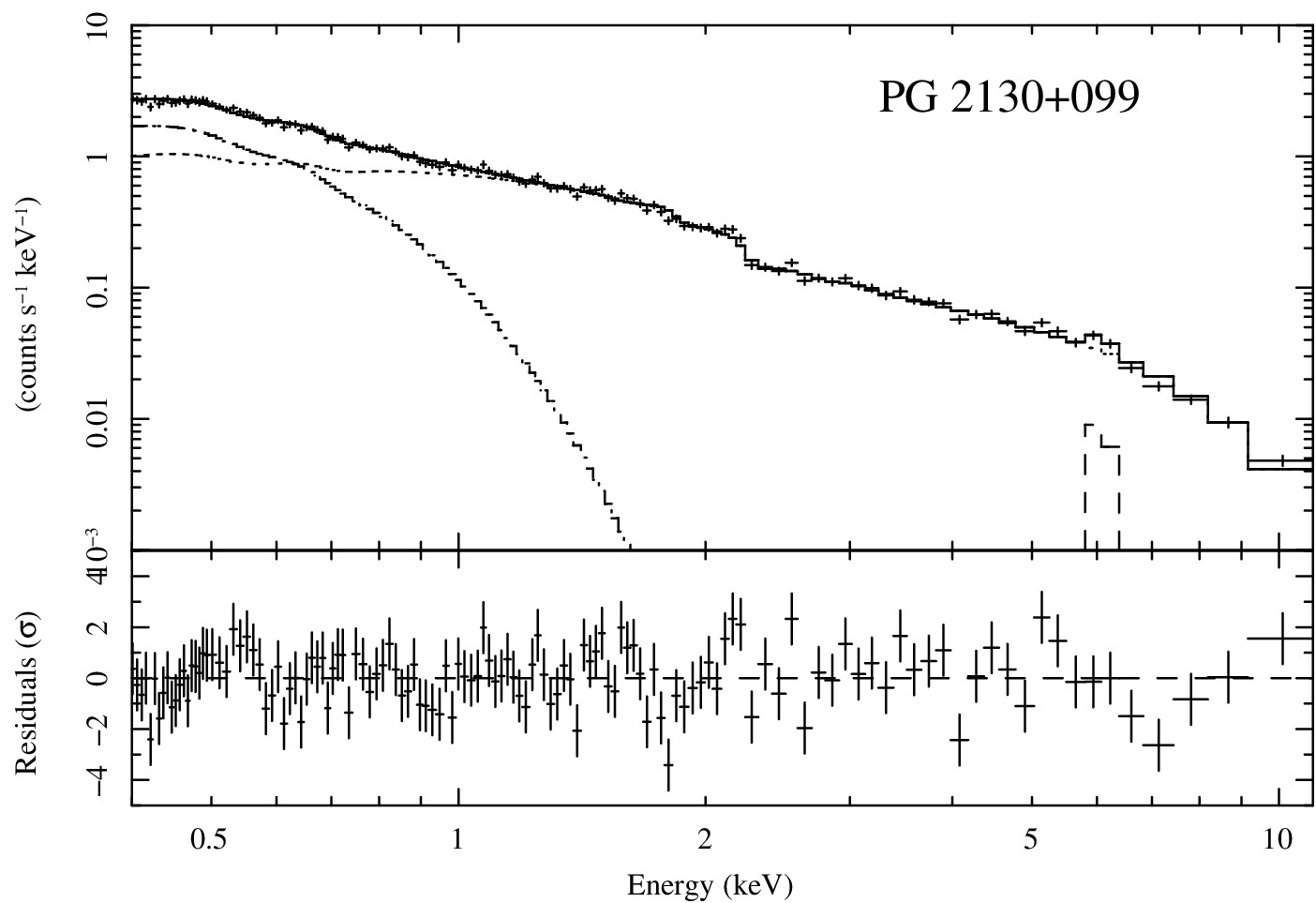}
\includegraphics[angle=-90,scale=0.28]{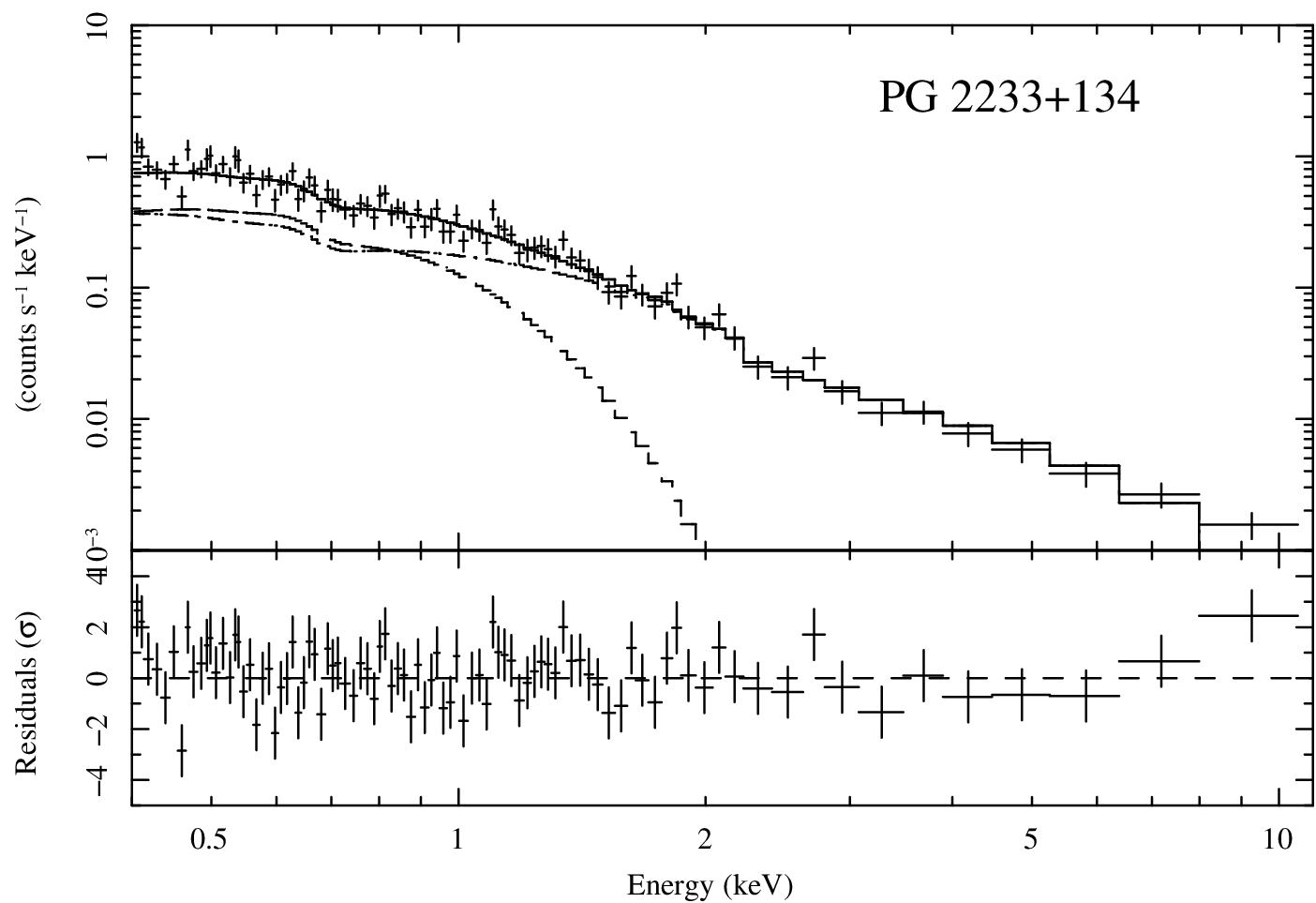}
\caption{
Spectra of PG quasars newly analyzed in this paper.
}
\label{fig: other quasars}
\end{center}
\end{figure}


\end{document}